# Eliminating Delocalization Error to Improve Heterogeneous Catalysis Predictions with Molecular DFT+U

Akash Bajaj[1,2] and Heather J. Kulik[1,*]

[1]*Department of Chemical Engineering, Massachusetts Institute of Technology, Cambridge, MA 02139*

[2]*Department of Materials Science and Engineering, Massachusetts Institute of Technology, Cambridge, MA 02139*

ABSTRACT: Approximate semi-local density functional theory (DFT) is known to underestimate surface formation energies yet paradoxically overbind adsorbates on catalytic transition-metal oxide surfaces due to delocalization error. The low-cost DFT+U approach only improves surface formation energies for early transition-metal oxides or adsorption energies for late transition-metal oxides. In this work, we demonstrate that this inefficacy arises due to the conventional usage of metal-centered atomic orbitals as projectors within DFT+U. We analyze electron density rearrangement during surface formation and O atom adsorption on rutile transition-metal oxides to highlight that a standard DFT+U correction fails to tune properties when the corresponding density rearrangement is highly delocalized across both metal and oxygen sites. To improve both surface properties simultaneously while retaining the simplicity of a single-site DFT+U correction, we systematically construct multi-atom-centered molecular-orbital-like projectors for DFT+U. We demonstrate this molecular DFT+U approach for tuning adsorption energies and surface formation energies of minimal two-dimensional models of representative early (i.e., $TiO_2$) and late (i.e., $PtO_2$) transition-metal oxides. Molecular DFT+U simultaneously corrects adsorption energies and surface formation energies of multi-layer models of rutile $TiO_2(110)$ and $PtO_2(110)$ to resolve the paradoxical description of surface stability and surface reactivity of semi-local DFT.

# 1. Introduction

Density functional theory (DFT) using approximate exchange-correlation (xc) functionals remains the primary quantum mechanical method of choice in computational modeling of heterogeneous catalysis[1-6]. However, semi-local xc approximations exhibit both one[7-10] and many-electron[9-12] self-interaction error (SIE), commonly referred to as delocalization error[13-15] (DE). DE leads to an erroneous prediction of dissociation energies[9,16-18], reaction barrier heights[19-22], adsorption energies[20,23-25] or adsorption site preferences[26-30] and surface energies[31-33] within semi-local DFT, properties that are crucial for an accurate modeling of heterogeneous catalysis. This has given rise to the development and application of both improved semi-local-DFT-based xc functionals targeted towards surface modeling[34-40] and approaches that attempt to reduce DE itself[41-43].

One widely adopted approach for reducing DE is the use of hybrid xc functionals incorporating some amount of one-SIE-free exact Hartree–Fock (HF) exchange[43-46]. Hybrid functionals can improve the prediction of dissociation or adsorption[47-52] energies, of reaction barrier heights[53-55] and of adsorption site preferences[56-58] (e.g., CO on Pt(111)). However, hybrid functionals simultaneously incur a larger computational cost relative to semi-local DFT[59-61] within plane-wave implementations, making them cost-prohibitive for high-throughput studies for the design and discovery of novel heterogeneous catalysts.

A commonly used alternative for reducing DE at the low cost of semi-local DFT is the DFT+U correction scheme[62-66]. In its simplified form[64,65], the DFT+U energetic correction:

$$E^{\mathrm{DFT+U}} = E^{\mathrm{DFT}} + \frac{1}{2}\sum_{I,nl}\sum_{\sigma} U_{nl}^{I}\left[\mathrm{Tr}\{\mathbf{n}_{nl}^{I,\sigma}(\mathbf{1} - \mathbf{n}_{nl}^{I,\sigma})\}\right], \qquad (1)$$

is determined by the $U$ parameter and the occupation matrices, $\mathbf{n}$, of the subshell $nl$ and spin $\sigma$ at atomic site $I$:

$$\{\mathbf{n}_{nl}^{I,\sigma}\}_{mm'} = \Sigma_{k,v} \langle \psi_{k,v}|\phi_{m'}^{I}\rangle \langle \phi_{m}^{I}|\psi_{k,v}\rangle, \quad (2)$$

obtained using the projections of extended states, $\psi_{k,v}$, on atomic orbitals (AOs), $\phi^I$. In typical applications of DFT+U, these projection AOs, $\phi^I$, are the $d$ AOs of transition-metal sites that are particularly prone to DE due to their well-localized $d$ electrons[66]. Hence, the DFT+U scheme has been routinely employed for an improved description of the structural, electronic and catalytic properties of transition-metal-containing surfaces[67-74].

In heterogeneous catalysis modeling, a severe challenge for semi-local DFT that necessitates the use of such DE reduction approaches is a simultaneously accurate description of the stability and reactivity of surfaces[75-78]. Specifically, semi-local DFT describes surfaces containing transition metals to be too stable by underestimating their surface formation energy[31,75,76,79]:

$$E_\sigma = \frac{E_{\text{slab}} - NE_{\text{bulk}}}{2A}, \quad (3)$$

where $E_{\text{slab}}$ and $E_{\text{bulk}}$ denote the energies of the surface and the bulk models respectively, $A$ is the surface area and $N$ is the ratio of the number of atoms in the surface model to those in the bulk model. At the same time, semi-local DFT paradoxically binds adsorbates on these surfaces too strongly.[23,24,75,76]

Although both approaches reduce the DE, hybrid functionals and DFT+U are known to correct semi-local DFT errors in transition-metal-containing systems in a divergent fashion[76,80-82]. In previous work[76], this divergence was demonstrated in the modeling of surface stability and surface reactivity of transition-metal dioxides that are promising catalysts for the oxygen evolution reaction during electrochemical water splitting[83-90]. It was observed that tuning the HF exchange fraction within hybrid DFT improved both surface energy, $E_\sigma$, and surface reactivity (i.e., $\Delta E_O$) in these systems with respect to correlated wavefunction theory references. $\Delta E_O$ is computed using the O atom adsorption energy:

$$\Delta E_O = E(O^*) - E(*) - \{E(H_2O) - E(H_2)\}, \tag{4}$$

where $E(*)$ denotes the energy of the pristine surface and $E(O^*)$ denotes the energy of the surface with the O atom adsorbate. In contrast, tuning the $U$ parameter within DFT+U only improved either the description of surface stability or surface reactivity of transition-metal oxides depending on the early (e.g., Ti and V) versus late (e.g., Ir and Pt) nature of the metal.

The inability to tune both surface properties with standard DFT+U may not be a failure of the functional form of the +$U$ correction, which should never worsen the DE[91]. Nevertheless, the standard approach has been observed to be inefficient in reducing the DE in some transition-metal-containing systems[91-95]. Pathological examples for DFT+U have been demonstrated in molecular transition-metal complexes where there is strong metal–ligand hybridization (e.g., $Mn(CO)_6$), near-degenerate outer-valence electronic structure due to weak-field ligands (e.g., $Fe(H_2O)_6$), or for metal centers with a closed-shell electronic configuration[91,92]. Although the standard DFT+U approach is frequently employed for an improved prediction of electronic properties of transition-metal oxides, it has been noted to have limited effect on materials with a closed-shell electronic structure[93,95,96]. Characteristics associated with delocalization error for transition-metal oxides also gives rise to limitations while using DFT+U to model their structural[33,97], electronic[98,99] and catalytic properties[100,101].

Small effects on DE and related properties with application of moderate values of $U$ in DFT+U may be resolved by the application of $U$ on the outer-valence metal $s$ AOs[93,100,102] and on the AOs of the metal-connecting atoms[33,96,97,103,104]. However, the choice of appropriate AOs on which the $U$ needs to be applied requires an inspection of the electronic structure of the system of interest and further requires the determination and tuning of multiple $U$ parameters. We recently proposed an alternative approach for transition-metal complexes where a single $U$ parameter applied using systematically selected frontier molecular orbitals (MOs) as the projection orbitals, $\phi^I$, reduces DE much more efficiently, especially for the pathological cases[92]. Similar approaches have also been found to improve adsorption site preferences on transition-metal surfaces[28,105] and electronic properties of transition-metal oxides and their surfaces[106-108].

Though the low-cost approach of molecular DFT+U has found success in modeling electronic properties of transition metal systems, its potential as a systematic low-cost approach for simultaneously improving the paradoxical description of surface stability and reactivity of transition-metal oxides has yet to be determined. In this work, we demonstrate the use of the low-cost molecular DFT+U approach for simultaneously improving $E_\sigma$ and $\Delta E_O$ of rutile transition-metal oxides with efficiency comparable to that of higher computational cost hybrid xc functionals. We establish a systematic protocol for constructing chemically meaningful multi-atom-centered projectors that capture significant metal–oxygen hybridization changes during surface formation and O atom adsorption. Using this procedure, we highlight how the molecular DFT+U approach improves the semi-local DFT description of both surface reactivity and surface stability across early and late transition-metal oxides while bypassing the larger computational cost incurred by hybrid functionals. The rest of this article is outlined as follows. In Section 2, we provide the computational details of the calculations employed in this work. In Section 3, we analyze density redistribution during surface formation and O atom adsorption for early and late rutile transition-metal oxide surfaces and demonstrate the use of a molecular DFT+U approach on representative two-dimensional and multi-layer rutile transition-metal oxide surfaces. Finally, in Section 4, we provide our conclusions.

## 2. Computational Details.

*Structure Preparation.* All initial structures were built using the Atomic Simulation Environment (ASE)[109] toolkit v3.16.2. Bulk structures of the rutile-type tetragonal transition-metal dioxides, $MO_2$ (M = Ti, V, Ru, Rh, Ir, Pt), were generated using the experimental lattice parameters obtained from the Inorganic Crystal Structure Database (ICSD)[110] (Supporting Information Table S1). Structures for all (110) surface models were generated using the corresponding DFT-relaxed bulk lattice parameters at each level of theory and consisted of 15 Å of vacuum. Three-dimensional (3D) slabs were modeled using 2 × 1 unit cells that consisted of four trilayers (i.e., O–M–O repeats). The two-dimensional (2D) layers were modeled with the smallest unit cell to capture all unique metal sites: a 1 × 1 unit cell for surface energies in $PtO_2$ and a 2 × 1 unit cell for adsorption energies in $TiO_2$. Initial bond lengths between the metal adsorption site and the adsorbate O atom were always set to expected distances corresponding to chemisorption (i.e., < 2.0 Å, Supporting Information Table S2 and Figures S1–S2).

*DFT Calculations.* DFT and DFT+U calculations[62-64,66] with standard atomic projectors on the $d$ states were carried out with the plane-wave periodic boundary condition code Quantum-ESPRESSO[111] v5.1, using the PBE[112] semi-local generalized gradient approximation (GGA) to the exchange-correlation (xc) functional. The Hubbard $U$ was employed in increments of 1 eV (e.g., from 0 to 10 eV), as indicated throughout the text. Ultrasoft pseudopotentials[113,114] (USPPs) were employed throughout, which were obtained from the Quantum-ESPRESSO website[115] (Supporting Information Table S3). Plane-wave cutoffs for all calculations were 35 Ry for the wavefunction and 350 Ry for the charge density[76]. The default convergence threshold of $1\times10^{-6}$

Ry for the self-consistent field (SCF) energy error was used. To aid SCF convergence, an electronic temperature of 0.005 Hartree was applied and the mixing factor was reduced to 0.4 from its default value of 0.7.

Geometry optimizations were carried out using the BFGS quasi-Newton algorithm[116-120] with default convergence thresholds of $1\times10^{-3}$ Ry/bohr for the maximum residual force and $1\times10^{-4}$ Ry for the change in energy. Lattice parameters and atomic positions of all bulk $MO_2$ crystals were optimized using $12 \times 12 \times 12$ Monkhorst–Pack $k$-point grids (Supporting Information Table S1). Atomic positions were optimized for the 3D slab models using a smaller $4 \times 4 \times 1$ Monkhorst–Pack $k$-point grid. Only the outermost trilayers were optimized for the pristine 3D slab models and only the two topmost trilayers and the adsorbate atom were optimized for the decorated 3D slab models. All atomic positions were optimized for the 2D slab models using a $4 \times 9 \times 1$ Monkhorst–Pack $k$-point grid for the $PtO_2$ layer and $4 \times 4 \times 1$ Monkhorst–Pack $k$-point grid for the $TiO_2$ layer.

*Wannier Functions.* Plane-wave eigenstates $|\psi_{k,v}\rangle$ were converted into real-space Wannier functions[121] using the pmw.x utility available with the Quantum-ESPRESSO package for use as MO projectors in DFT+U (Supporting Information Text S1). All eigenstates were generated at the PBE level of theory (Supporting Information Table S4). Relevant eigenstates were selected (see Sec. 3b and Sec. 3c) by specifying a contiguous range for the band index, $v$, using the "first_band" and "last_band" keywords (Supporting Information Text S1). The MOP DFT+U properties were evaluated from single-point energies using the PBE optimized geometry with $U$ values applied to the MOPs ranging from 0 eV (i.e., PBE) to up to 5 eV in 1 eV increments. This step was necessary because forces are not available for DFT+U with the Wannier function-based MO basis, but comparisons for AO-based DFT+U where optimized structures could be used indicated a limited effect of geometry changes (Supporting Information Table S5).

*Density and Partial Charge Analysis.* Electron density cube files were obtained using the pp.x postprocessing utility of Quantum-ESPRESSO. The grid resolutions were manually specified to ensure equivalent grid point spacing for computing density differences (Supporting Information Table S6). Extraction of the density in the (110) plane was carried out using an in-house BASH script, and the density difference was computed using an in-house MATLAB script. All structures and three-dimensional densities were visualized using VESTA[122] v3.3.9. Löwdin partial charges[123] were computed using the projwfc.x utility of Quantum-ESPRESSO.

## 3. Results and Discussion.

### 3a. Density Delocalization Effects on Surface and Adsorption Energies.

Surface formation energies ($E_\sigma$, see eq. (3)) of rutile dioxides have previously been observed to be too low with semi-local DFT, as has the tendency of hybrids and DFT+U to correct this error to differing degrees.[76] Specifically, the sensitivity of $E_\sigma$ for the $MO_2$(110) surface to the $U$ value in DFT+U, $S_U(E_\sigma)$, was similar in magnitude to that from Hartree–Fock (HF) exchange tuning in hybrids, $S_{HF}(E_\sigma)$ only for early transition-metal dioxides (e.g., Ti and V, Supporting Information Table S7). In contrast, later transition metals (TMs) (e.g., Rh and Pt) had very low $S_U(E_\sigma)$ values for the (110) plane in comparison to equivalent $S_{HF}(E_\sigma)$ values (Supporting Information Table S7). One reason why $S_U(E_\sigma)$ could be so low is that there might be significant changes in metal–oxygen hybridization during surface formation that are not detected with a Hubbard U projector applied only to the $d$ states of the metal atoms of later TMs. To understand if that is the reason for $d$-filling dependence of $S_U(E_\sigma)$, we analyzed the electron density variation at both the metal and oxygen atom sites for early and later TM surface formation.

We first computed the electron density difference in the MO$_2$(110) plane by substracting a 2D density slice of the 3D grid between the bulk rutile oxide and the pristine slab (Figure 1 and Supporting Information Figures S3–S8). The density redistribution that occurs in this plane is qualitatively different for earlier (e.g., Ti, V) and later (e.g., Rh, Pt) TMs even though they all have the same qualitative MO$_2$ geometric structures. For the earlier TMs, the density change upon surface cleavage is fairly localized on the surface metal site (Figure 1 and Supporting Information Figures S3–S4). This observation, combined with their relatively high $S_U(E_\sigma)$ values[76], suggests that a standard DFT+U approach (i.e., with metal-centered atomic orbital projectors) adequately detects an increase in delocalization during surface formation. For the later TMs, electron density redistribution is balanced between surface metal and oxygen sites (Figure 1 and Supporting Information Figures S5–S8). Although one might expect estimates of density redistribution to be sensitive to the choice of plane[124], modest variations (ca. ± 0.08 Å) in choice of the plane position for analysis do not alter our observations (Supporting Information Figures S3–S8). Taken together with a low $S_U(E_\sigma)$ value for the representative PtO$_2$ case, this observation indicates that the rehybridization involving both the metal and oxygen that occurs for later TMs cannot be adequately detected within a standard DFT+U approach.

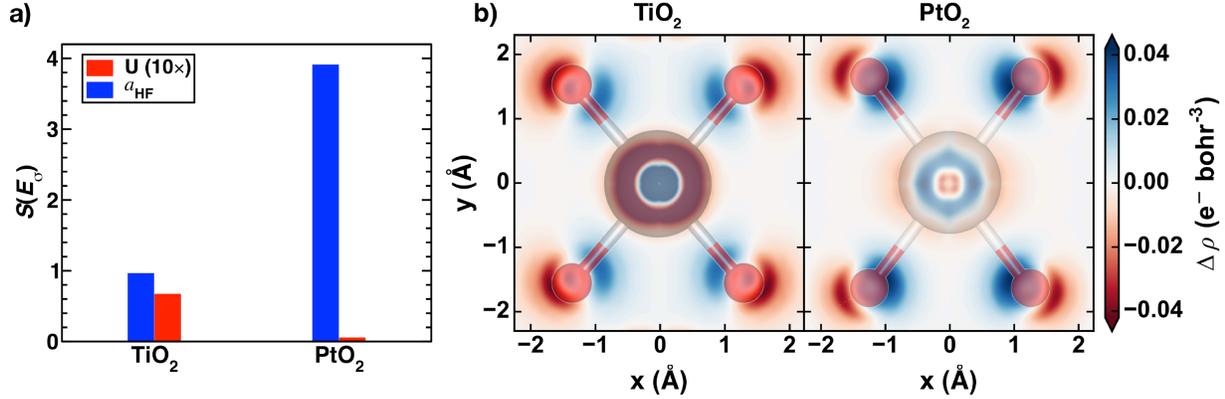

**Figure 1.** (a) Linearized sensitivity of the surface formation energy, $S(E_\sigma)$, for the (110) planes of rutile $TiO_2$ and $PtO_2$. Sensitivities were computed with respect to the $U$ value (red bars) in units of eV/(110) u.a. per eV of $U$ (and multiplied by 10) or were obtained from Ref. [76] (in eV/(110) u.a. per HFX) where it was computed with respect to a change in the Hartree–Fock (HF) exchange fraction, $a_{HF}$, from 0 to 1 i.e., 1 HFX unit (blue bars). (b) Density difference, $\Delta\rho$, (in e$^-$/bohr$^3$) between the pristine slab and the bulk rutile models of $TiO_2$ (left) and $PtO_2$ (right) computed on the (110) plane. Red indicates density loss and blue indicates density addition evaluated after the surface has been cleaved. Translucent models of the metal atom (gray sphere) and the four coordinating oxygen atoms (red spheres) are overlaid.

We quantified the relative contribution of the in-plane density redistribution at the metal and the oxygen sites by integrating the absolute density difference, $|\Delta\rho|$, on the (110) plane within a 2D circular region centered on the two sites (Supporting Information Figure S9). For earlier TM oxides, the metal-local integrated planar density contribution was more than twice as large as compared to that for late TM oxides. This early TM metal-local contribution was so large that it was comparable to those from all the four coordinating oxygen atoms combined (i.e., roughly 50% of the total density difference) (Supporting Information Table S8). For the later TM oxides, the metal-local contribution to the planar density redistribution was comparatively less (i.e., 25–28%). This $d$-filling-specific trend of the integrated density difference is independent of the definition of the radius of the oxygen atom (Supporting Information Figure S9). Hence, significant oxygen-centered contributions to density changes for later TM oxides suggest that a multi-atom-centered projector basis will be required to obtain higher DFT+U surface formation energy sensitivities.

In comparison to surface formation energies, DFT+U and hybrid tuning on oxygen-atom adsorption energies ($\Delta E_O$, see eq. (4)) of rutile dioxides were shown to exhibit opposite trends.[76] Specifically, the sensitivity of $\Delta E_O$ with DFT+U (i.e., $S_U(\Delta E_O)$) was higher for later TMs (e.g., Rh and Pt) and negligible for early TMs (e.g., Ti and V), while Hartree–Fock (HF) exchange tuning (i.e., $S_{HF}(\Delta E_O)$) is large for all rutile dioxides (Supporting Information Table S9). Because we correlated changes in surface metal–oxygen hybridization to $S_U(E_\sigma)$ values, we expect that the nature of changes in surface metal–oxygen (M–O) and metal–adsorbate-oxygen (M–O*) hybridization should rationalize relative $S_U(\Delta E_O)$ values.

To identify trends with $d$ filling, we computed the electron density difference over the entire $MO_2$(110) slab model after adsorption (i.e., of a ½ monolayer of O atoms) and analyzed the three-dimensional density difference, $\Delta\rho$, near the adsorption site as a proxy for changes in M–O hybridization (Figure 2 and Supporting Information Figure S10). The density redistribution during adsorption exhibits qualitative differences between early and late TMs (Figure 2 and Supporting Information Figure S10). For the late TM oxide $PtO_2$, the density redistribution within the (110) plane upon adsorption is mostly localized on the metal and secondarily on the adsorbing oxygen (Figure 2). This large $\Delta\rho$ local to the metal corresponds to changes in the electron occupations of the metal $d$ atomic orbitals[76] upon adsorption, meaning that a standard atomic basis for DFT+U can be expected to produce substantial $S_U(\Delta E_O)$ values for the later TMs (Figure 2 and Supporting Information Table S9). In contrast, for the early TM oxide $TiO_2$, the density redistribution is highly delocalized on the (110) plane and away from the surface metal site. This delocalization of the density redistribution decreases from early to late TMs (V to Ru or Rh in Supporting Information Figure S10). Such a significant change in the M–O hybridization away from the metal site corresponds to the much lower $S_U(\Delta E_O)$ values for early TMs with a $U$ applied only on the metal-local $d$ states.

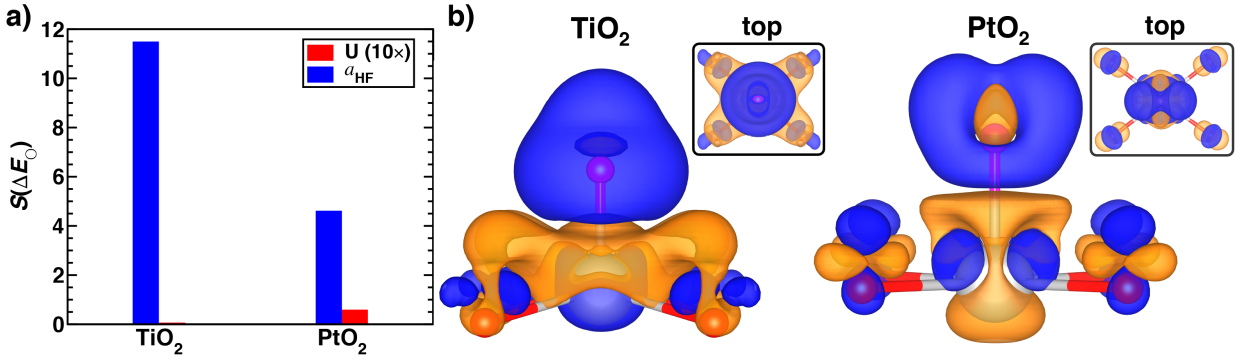

**Figure 2.** (a) Linearized sensitivity of the adsorption energy, $S(\Delta E_O)$, for the adsorption of 0.5 monolayer of oxygen atoms (O*) on the (110) plane of rutile $TiO_2$ and $PtO_2$. Sensitivities were computed with respect to the $U$ value (red bars) in units of eV/eV of $U$ (and multiplied by 10) or were obtained from Ref. [76] (in eV/HFX) where it was computed with respect to a change in the Hartree–Fock (HF) exchange fraction, $a_{HF}$, from 0 to 1 i.e., 1 HFX unit (blue bars). (b) Isosurfaces for the density difference (isovalue: 0.003 e⁻/bohr³) between O*-adsorbed and pristine (110) surfaces of $TiO_2$ (left) and $PtO_2$ (right). The metal adsorption site (gray spheres), the four in-plane coordinating O atoms (red spheres), and the O* adsorbate (pink spheres) are shown in the front view with the top view depicted in the inset. Orange indicates density loss and blue indicates density gain after O* adsorption.

We further quantified the influence of M–O* hybridization changes on $S_U(\Delta E_O)$ by computing the maximum absolute $\Delta\rho$ value along the M–O* bond (Supporting Information Table S10). The maximum absolute $\Delta\rho$ is higher for the early TMs (> 0.1 e⁻/Å³) than for the later TMs (< 0.1 e⁻/Å³). This larger bond-centered density difference in early TM dioxides highlights why a metal-centered DFT+U approach leads to low values of $S_U(\Delta E_O)$ for early TMs[76] in comparison to late TMs (Supporting Information Table S9). Taken together with differences in the in-plane rehybridization, we expect that the standard DFT+U approach with a $U$ applied on metal $d$ states will have a limited effect on $\Delta E_O$. Indeed, $S_U$ values for $\Delta E_O$ are consistently below the $S_{HF}$ values for all TMs in comparison to trends in $E_\sigma$ evaluations, motivating an alternative to atomic $d$ orbitals for the states considered in DFT+U corrections of adsorption energies (Supporting Information Tables S7 and S9). A possible approach that has been pursued is to apply a $U$ on both the metal $d$ and oxygen $p$ states[33,96,125,126], but this requires the selection of two appropriate $U$ values, which are known to be system- and property-dependent. As an alternative, we next investigate the possibility of a single $U$ value in DFT+U applied over a multi-atom-centered basis to obtain DFT+U sensitivities comparable to those obtained from hybrids simultaneously for $E_\sigma$ and $\Delta E_O$ with both early and late TMs.

**3b. Studying 2D MO$_2$ with DFT+U using Molecular Projectors.**

In DFT+U, the functional form of the correction consists of a $U$ parameter multiplied by the fractionality[127], Tr[**n**(1-**n**)] (see eq. (1)). When DFT+U is applied in surface science, values of **n** are typically computed by projecting the extended states onto the metal-centered $d$ atomic orbitals as projectors (AOPs) (see eq. (2)). To a first-order approximation, the expected value of $S_U(E_\sigma)$ can be predicted by the metal-local fractionality difference[76] between the surface and the bulk models i.e., $\Delta$Tr[**n**(1-**n**)] computed with PBE (i.e., $U = 0$ eV). Thus, adjusting the projectors to molecular orbital projectors (MOPs) that increase the $\Delta$Tr[**n**(1-**n**)] between surface and bulk can be expected to improve sensitivities by more adequately reflecting the density redistribution that occurs upon surface cleavage. To determine if improvement in sensitivities is feasible with MOPs, we evaluated the effect of projector choice on $S_U(E_\sigma)$ for a 2D model of the representative later TM PtO$_2$ for which surface formation energies with AOPs in the 3D system were insensitive to DFT+U.

The choice to analyze a simplified 2D model consisting only of the single top layer in rutile PtO$_2$(110) simplifies the evaluation of Wannier functions by requiring consideration of only two unique Pt sites (Figure 3 and Supporting Information Figure S1). To confirm the suitability of this model system, we computed the AOP $\Delta$Tr[**n**(1-**n**)] and confirmed it to be negligible, with the exfoliation energy (i.e., $E_\sigma$ for this single layer) changing by < 0.1 eV from $U$ = 0 to 5 eV (Figure 3 and Supporting Information Table S11). Because only one Pt site, $Pt_s$, has Pt–O bonds that break during exfoliation and becomes undercoordinated, we replace the AOPs only at $Pt_s$ with MOPs and leave the AOPs unchanged for the other Pt site, $Pt_b$ (Figure 3 and Supporting Information Figure S1). As a result, the 2D $PtO_2$ $S_U(E_\sigma)$ should be determined only by the $\Delta Tr[\mathbf{n(1-n)}]$ for the new projectors on the $Pt_s$ site. This strategy can be generalized, for example to surface properties of 3D rutile $MO_2$ systems by retaining AOPs for multiple bulk-like (e.g., subsurface sites in a slab) metal sites (see Sec. 3c).

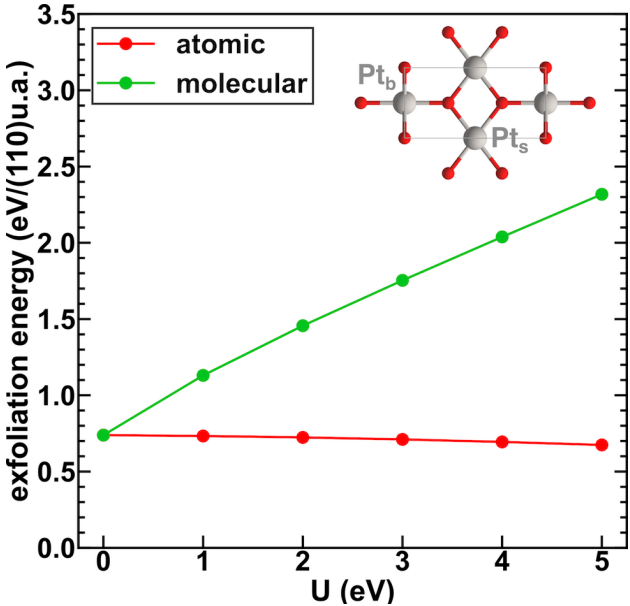

**Figure 3.** Exfoliation energy of 2D $PtO_2$ (in eV/(110)u.a.) computed using different projectors within DFT+U at $U$ values ranging from 0 to 5 eV. Energies were computed using atomic projectors (red) at both Pt sites or replacing only the atomic projectors at the surface undercoordinated Pt site, $Pt_s$, with the fractionality-selected best molecular projectors constructed using the Wannier localization scheme (green). The top view of the unit cell of 2D $PtO_2$ is shown in the inset with the $Pt_s$ site annotated.

To construct multi-atom-centered projectors using $\Delta Tr[\mathbf{n(1-n)}]$ as the figure of merit, we use the approximate Wannier function (WF) localization scheme implemented in Quantum-ESPRESSO[111] as in prior work[92] (Supporting Information Text S1). We must select a set of contiguous plane-wave states $\{|\psi_{\mathbf{k},v}\rangle\}$, where **k** denotes the *k*-point and *v* denotes the band index, and we project them onto the AOPs being replaced (here, the Pt$_s$ 5*d* AOs). These states are then transformed into five real-space WF projectors, analogous to molecular orbitals as projectors (MOPs) in isolated transition-metal complexes[92]. (Supporting Information Text S1). For both bulk and 2D PtO$_2$, we select all combinations of five contiguous states from a total of 50 possible states both above and below the Fermi level (Supporting Information Table S4). We thus construct 46 sets of MO-like WF projectors each for bulk and 2D PtO$_2$ i.e., leading to 2116 possible (i.e., 46 × 46) pairs. Several criteria exist for selecting a good set of MO projectors to increase the sensitivity of the exfoliation energy to Hubbard U values. Because surface states are more likely to be present near the Fermi level and have partial occupancies, we require a non-zero value of Tr[**n(1-n)**] (i.e., partial occupancy) of the MO projectors in both bulk and 2D PtO$_2$ for the Pt$_s$ site (Figure 4). Additionally, we select the MOP pair having the maximum positive ΔTr[**n(1-n)**] between the bulk and the surface because the corresponding states are likely undergoing the most significant change in hybridization during exfoliation (Figure 4). Using these two criteria, we identify a pair of MO projectors for bulk and 2D PtO$_2$ to maximally increase the $S_U(E_\sigma)$ of 2D PtO$_2$ (Figure 4). Although we selected a single pair to test our approach, a number of other pairs around the Fermi level would have also satisfied both criteria (Figure 4).

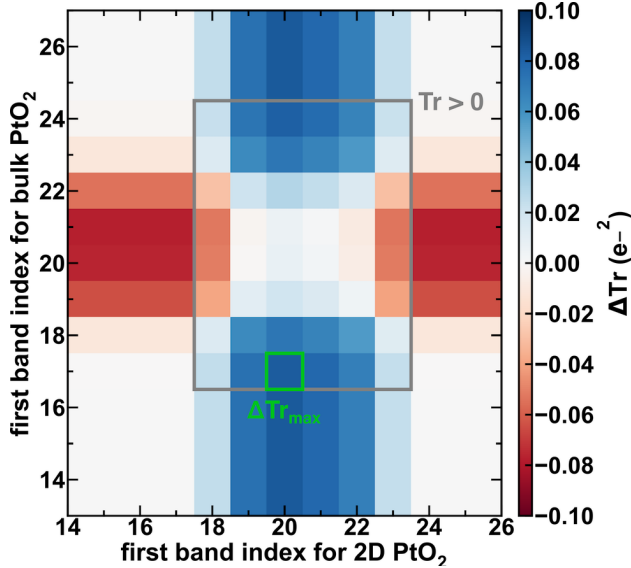

**Figure 4.** Fractionality difference per orbital, $\Delta \text{Tr}[\mathbf{n}(1-\mathbf{n})]$ (in $e^{-2}$), between 2D $PtO_2$ (110) and bulk rutile $PtO_2$ shown for representative molecular projectors constructed using the Wannier localization scheme for both systems. The index of the first band selected for the Wannier localization for 2D $PtO_2$ is labeled on the x-axis and the index of the first band selected for the Wannier localization for bulk $PtO_2$ is labeled on the y-axis. The projector pair having the maximum $\Delta \text{Tr}[\mathbf{n}(1-\mathbf{n})]$ is outlined with a green square, and those having a positive $\text{Tr}[\mathbf{n}(1-\mathbf{n})]$ on $Pt_s$ are outlined with a gray rectangle.

To gain physical intuition for the numerically selected best MOPs, we analyze the real-space (i.e., Γ-point) projected density of states (PDOS) and densities of the selected MOPs for 2D $PtO_2$ (Figure 5). The selected MO states reside within 1.0–1.5 eV of the Fermi level (Figure 5). As could be expected, the MOs have contributions from both Pt($5d$) and O($2p$) AOs, but do not have a majority contribution from Pt($5d$), unlike the standard AOPs (Supporting Information Table S12). Another distinct feature of all the MOs is the prominent contribution (>15%) from the O($2p_z$) AOs where the z-axis is perpendicular to the (110) plane (Figure 5). We expect O($2p_z$) AO contributions to be important because their out-of-plane electron density is likely to get redistributed during exfoliation due to changes in the local bonding environment. Indeed, no other set of five contiguous states were found to have as large O($2p_z$) AO contributions as the ones selected by our figure of merit (Supporting Information Table S12).

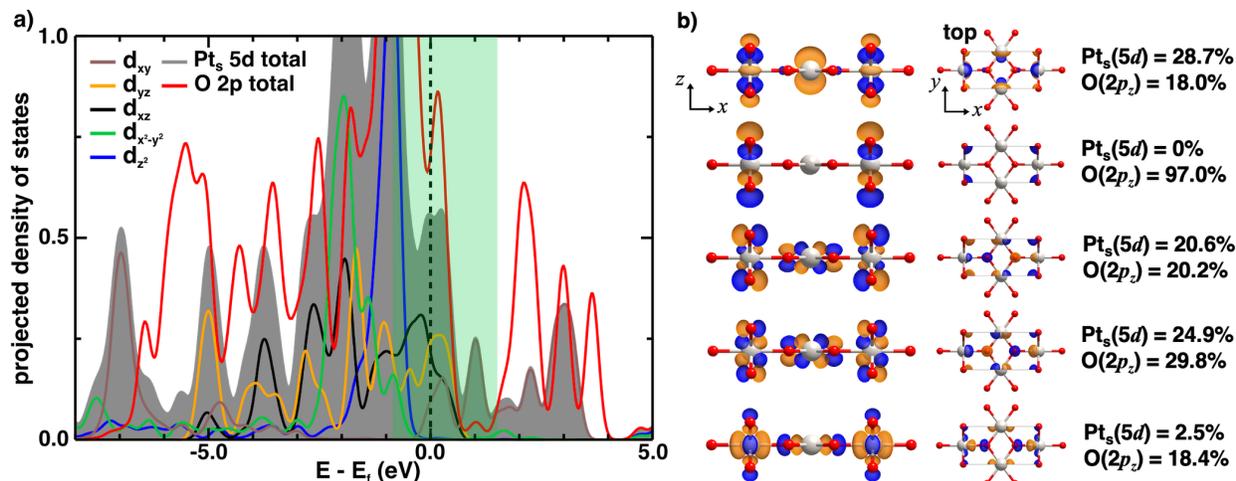

**Figure 5.** (a) Γ-point projected density of states (PDOS) of all $Pt_s(5d)$ atomic orbitals (gray shaded region) and all nearest-neighbor $O(2p)$ atomic orbitals (red solid line) in 2D $PtO_2$ (110). PDOS for the $5d_{xy}$, $5d_{yz}$, $5d_{xz}$, $5d_{x^2-y^2}$ and $5d_{z^2}$ atomic orbitals are also indicated using brown, orange, black, green, and blue solid lines respectively. The vertical black dashed line indicates the Fermi level ($E_f$) while the green transparent region represents the states selected for constructing the molecular projectors within the Wannier localization scheme. (b) Γ-point density isosurfaces (isovalue = 0.005 e⁻/bohr³) of the five molecular projectors for 2D $PtO_2$. Blue indicates positive wavefunction phase and orange indicates negative wavefunction phase. States are ordered in increasing energy from top to bottom. Views along both the y-axis and the z-axis are shown with $Pt_s(5d)$ and $O(2p_z)$ contributions annotated. Pt and O atoms are shown as gray and red spheres respectively.

To validate the effect of MOPs on DFT+U energy corrections, we computed the exfoliation energy of 2D $PtO_2(110)$ (i.e., $E_\sigma$) with increasing $U$ values (Figure 3 and Supporting Information Table S11). Compared to the negligible change in $E_\sigma$ observed with AOPs, $E_\sigma$ obtained using MO projectors increases by more than 1.5 eV over a 5 eV increase in $U$, which corresponds to a twenty-fold increase in $S_U(E_\sigma)$. Clearly, the construction of MOPs using Wannier localization and fractionality-based metrics provides us with physically justified multi-atom-centered projectors having higher $S_U(E_\sigma)$ than possible with metal-centered AOPs. Moreover, the use of such a projector choice facilitates the tuning of $E_\sigma$ using a single $U$ parameter.

We next investigated how the use of MOPs influences early TM cases (e.g., $TiO_2$) where standard AOPs led to low sensitivities to applied U for oxygen adsorption, $S_U(\Delta E_O)$, and test our approach on a 2D monolayer. We again focus on the 2D model due to the smaller number of sites and bands over which we need to select and evaluate Wannier-based MOPs (Supporting Information Figure S2). We estimate the expected $S_U(\Delta E_O)$ with AOPs for this system from the fractionality difference, $\Delta\mathrm{Tr}[\mathbf{n}(1-\mathbf{n})]$ at $U = 0$ eV, between the surface with adsorbed O and the pristine surface[76]. As expected, the $\Delta\mathrm{Tr}[\mathbf{n}(1-\mathbf{n})]$ for O atom adsorption on 2D TiO$_2$(110) is negligible with DFT+U using AOPs (i.e., < 0.01 eV/eV of $U$, Figure 6 and Supporting Information Table S13).

We replace the AOPs with MOPs for the adsorption site, Ti$_s$, and treat the other two Ti sites in this 2D TiO$_2$ model (i.e., Ti$_b$ and Ti$_u$) with standard AOPs (Figure 6 and Supporting Information Figure S2). Similar to the case of PtO$_2$, we select all combinations of five contiguous states from a total of 75 possible bands both below and above the Fermi energy for the pristine or the O*-decorated 2D TiO$_2$ (Supporting Information Text S1 and Table S4). This combination leads to 5041 possible sets (i.e., 71 × 71) of pairs of MOPs because the band index need not be identical for the two (i.e., pristine and adsorbed) 2D TiO$_2$ models. We use the same two criteria we applied for $S_U(E_\sigma)$, i.e., $\mathrm{Tr}[\mathbf{n}(1-\mathbf{n})] > 0$ for the MOPs and a maximum value of $\Delta\mathrm{Tr}[\mathbf{n}(1-\mathbf{n})]$, to identify a single best pair of MO projectors (Supporting Information Figure S11).

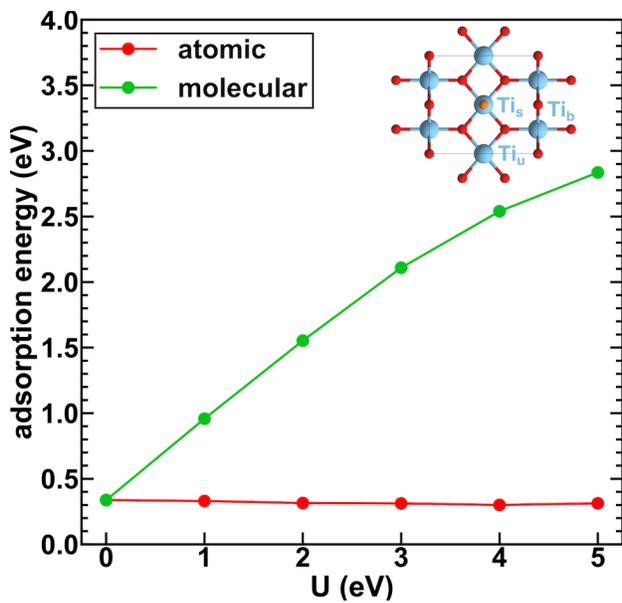

**Figure 6.** Adsorption energy (in eV) for a ½ monolayer of O on 2D $TiO_2$(110) computed using different projectors within DFT+U at $U$ values ranging from 0 to 5 eV. Energies were computed using atomic projectors (red) at all three Ti sites or replacing the atomic projectors only at the site of adsorption, $Ti_s$, with the fractionality-selected best molecular projectors constructed using the Wannier localization scheme (green). The top view of the unit cell of the O* adsorbate containing 2D $TiO_2$ is shown in the inset. Ti atoms are shown in blue, O atoms are shown in red, the adsorbate O* atom is shown in orange, and the $Ti_s$ site is annotated.

The fractionality-selected MO states lie within 1.5 eV of the Fermi level (i.e., both above and below) for the O*-decorated 2D $TiO_2$ whereas the corresponding MO states for the pristine $TiO_2$ layer all lie within 1.5 eV below its Fermi level (Figure 7 and Supporting Information Figure S12). The O(2$p$) AO contributions from surface oxygen atoms are > 40% for the MOPs in both models and significantly outweigh the $Ti_s$(3$d$) contributions (< 10%) in all MOPs (Figure 7 and Supporting Information Tables S14–S15). In particular, for the adsorbed model, the $TiO_2$ MOPs have a prominent (>35%) contribution from the O*(2$p$) states, which follows our expectations of what would be required to obtain positive $S_U(\Delta E_O)$ values. These MOPs corresponds to weak $\pi$* interactions, as indicated by a mixture of $Ti_s$(3$d_{xz}$) and $Ti_s$(3$d_{yz}$) AOs with O O*(2$p$) states. The MOPs corresponding to stronger Ti–O* $\sigma$ overlap lie well below the Fermi level and are fully occupied (i.e., Tr = 0), leading them not to be selected by fractionality analysis (Supporting Information Table S15). Furthermore, no other set of five contiguous states were found to have as large O*(2p) AO contributions (Supporting Information Table S15).

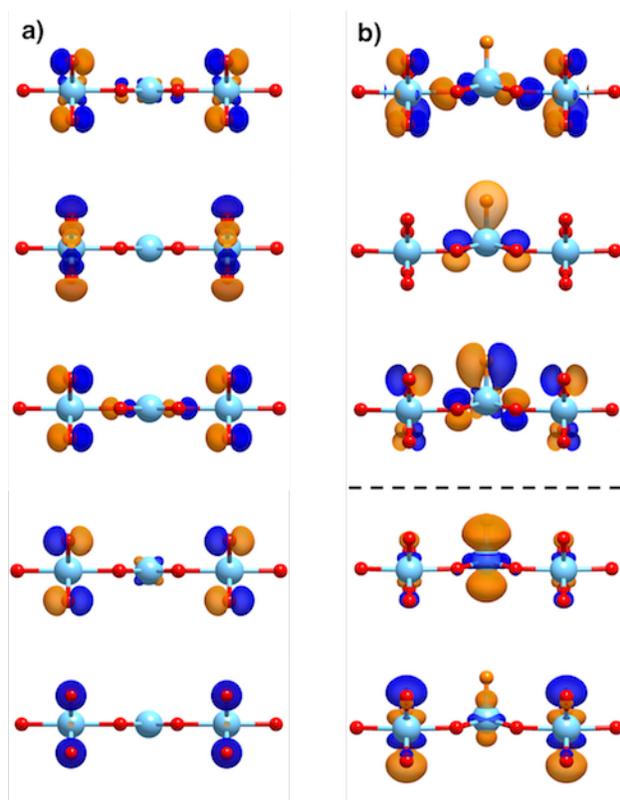

**Figure 7.** Γ-point density isosurfaces (isovalue = 0.002 e⁻/bohr³) of the five fractionality-selected best molecular projectors for (a) pristine 2D $TiO_2$ and (b) O*-decorated 2D $TiO_2$ shown in the side view. Blue indicates positive wavefunction phase and orange indicates negative wavefunction phase. States are ordered in increasing energy from top to bottom, where energies of all states for pristine 2D $TiO_2$ were lower than its Fermi level and only the states above the black horizontal dashed line for O*-decorated 2D $TiO_2$ had energies lower than its Fermi level. Ti, O and adsorbate O* atoms are shown as blue, red and orange spheres respectively.

Finally, we validated the suitability of these MOPs by computing the $\Delta E_O$ of 2D $TiO_2(110)$ with increasing $U$ values (Figure 6 and Supporting Information Table S13). For the fractionality-selected MOPs, the $\Delta E_O$ increases by 2.5 eV over a 5 eV increase in $U$, resulting in an $S_U(\Delta E_O)$ increase of a factor of more than 80 relative to the AOPs. While these selected MOPs can be expected to yield the largest sensitivities, they have asymmetric contributions of $Ti_s(3d_{xz})$ and $Ti_s(3d_{yz})$ AOs and thus break the degeneracy of these states. We therefore also considered an alternative, contiguous set of five MOPs by adjusting the selection until there was an equal contribution from both Ti$_s$(3$d_{xz}$) and Ti$_s$(3$d_{yz}$) AOs. While this set does not produce the maximum sensitivity (i.e., as judged by our $\Delta\text{Tr}[\mathbf{n}(1-\mathbf{n})]$ figure of merit), it still significantly exceeds that obtained with AOPs (Supporting Information Tables S13–S14). This alternate MOP selection results in an $S_U(\Delta E_O)$ higher by a factor of more than 70 relative to the AOPs. Thus, one could directly focus on constructing physical, symmetric MOPs based on intuition to achieve the same result of relatively high sensitivities, but it would be harder to automate. Overall, MOPs constructed using Wannier localization and fractionality-based metrics are able to yield higher $S_U(\Delta E_O)$ in comparison to to the use of AOPs.

### 3c. Simultaneous Tuning of Surface Formation and Adsorption Energies for MO$_2$ Slabs.

We then asked whether our MOP approach could simultaneously tune $S_U(E_\sigma)$ and $S_U(\Delta E_O)$ in 3D (i.e., slab) multi-layer models of rutile MO$_2$. For the later TMs, we return to the prototypical case of PtO$_2$, where tuning $U$ in the standard DFT+U approach destabilizes O atom adsorption (i.e., high $S_U(\Delta E_O)$) but has no effect on the surface formation energy (i.e., low $S_U(E_\sigma)$)[76]. We treat all distinct Pt sites uniquely and replace the projectors only at the Pt sites involved in Pt–O bond breaking in surface formation or those involved in Pt–O* adsorption (i.e., Pt$_u$ and Pt$_s$ sites) (Supporting Information Figure S1). For the single Pt$_s$ site in bulk rutile PtO$_2$, we replace the five Pt$_s$ $d$ AO projectors with five MOPs, leading to 46 possible combinations of five contiguous states from the total 50 available both above and below the Fermi level (Figure 8 and Supporting Information Table S4). For pristine and O*-decorated PtO$_2$(110) surface models that contain four such Pt sites (i.e., Pt$_s$ or Pt$_u$), we now need to replace a total of 20 Pt 5$d$ AOs with 20 MOPs (Supporting Information Figure S1). Thus, for each of the two surface models, we select all possible combinations of 20 contiguous states from the total 250 available to construct 231 sets of MOPs (Figure 8 and Supporting Information Table S4). After constructing the MOP sets, we determine the best possible pairs of sets that maximize the effect of Hubbard U on the properties of interest using the same criteria as before. Specifically, we compute Tr[**n(1-n)**] for each MOP set, require it to be positive, and compute ΔTr[**n(1-n)**] for each MOP pair relevant to both the $E_\sigma$ and $\Delta E_O$ properties (Figure 8). The $S_U(E_\sigma)$ and $S_U(\Delta E_O)$ are influenced by the difference (i.e., ΔTr[**n(1-n)**]) between the pristine and bulk models and between the decorated and pristine models respectively. Hence, if the dangling bonds or adsorbate-involved MOs are the most different between these two reference states, the ΔTr[**n(1-n)**], and thus the sensitivities, should be maximized for the MOPs (Figure 8). We observe that a large number of pairs satisfy these two criteria (Figure 8). Generally, higher-energy orbitals are preferred for the pristine slab in the case of the adsorption energy than for surface formation, but a range of 10–20 starting band numbers is suitable for both, whereas a narrower range of band indices is suitable for the smaller bulk system (Figure 8).

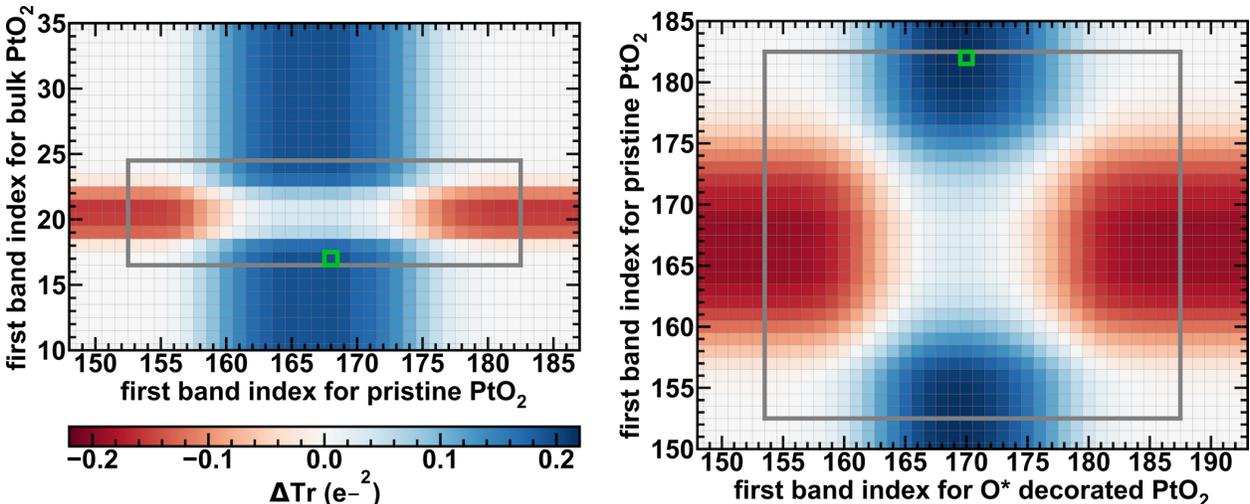

**Figure 8.** Fractionality difference per orbital, ΔTr[**n(1-n)**] (in e$^{-2}$), between representative molecular projectors constructed using the Wannier localization scheme of the pristine PtO$_2$(110) surface and bulk rutile PtO$_2$ (left) and of O*-decorated and pristine PtO$_2$(110) surfaces (right). Values on the x and y axes show the index of the first band selected for Wannier localization in the indicated system. The projector pair having the maximum $\Delta\text{Tr}[\mathbf{n}(1\text{-}\mathbf{n})]$ is outlined with a green square and those having a positive Tr are outlined with a gray rectangle.

As could be expected from our criteria, the best MOPs are within 3 eV of the Fermi level for all models (Supporting Information Figures S13–S15). For bulk $PtO_2$, the selected MOPs predominantly consist of O(2$p$) states and are thus distinct from the metal-centered AOPs in standard DFT+U (Supporting Information Figure S13). For both the pristine and O*-adsorbed $PtO_2$(110) surface models, most of the selected MOPs also have O(2$p$) contributions that are higher than or comparable to the Pt(5$d$) contributions (Supporting Information Tables S16–S17). Thus, projectors that discriminate the properties of the bulk from either the pristine or decorated slab must directly address differences in oxygen orbital occupations. For example, the Pt(5$d$) contributions to the pristine $PtO_2$(110) slab MOPs selected for $S_U(E_\sigma)$ predominantly consist of out-of-plane Pt(5$d_{z^2}$) AOs (Supporting Information Figure S14). For O*-decorated $PtO_2$(110) MOPs relevant to $S_U(\Delta E_O)$, a large contribution (ca. 40–55%) comes from the O*(2$p$) AO (Supporting Information Figure S15 and Table S17). Thus, while our MOPs are empirically selected to maximize fractionality differences, our criteria naturally led to selection of MOPs that reflect the chemical bonding changes between the states being compared.

Finally, we use the fractionality-selected best MOPs to evaluate $E_\sigma$ and $\Delta E_O$ for $PtO_2$ and the corresponding linearized DFT+U sensitivities (Figure 9). We observe that the use of the best selected MOPs achieves our goal of tuning both properties substantially more than the use of AOPs. In fact, the MOP DFT+U sensitivities are one to two orders of magnitude higher than the AOP DFT+U sensitivities, suggesting the possibility of using lower $U$ values for these best selected MOPs, which can have the advantage of improving calculation convergence in practice[66,128]. Most importantly, MOPs simultaneously tune $E_\sigma$ and $\Delta E_O$ for the late transition-metal oxide $PtO_2$. The fractionality-selected best MOPs for the pristine $PtO_2$(110) model are different for tuning $E_\sigma$ and $\Delta E_O$ (Figure 8 and Supporting Information Table S16). To reduce any ambiguity this introduces, we selected a compromise set of contiguous MOPs for the pristine PtO$_2$(110) slab that were near the Fermi level (i.e., Tr[**n(1-n)**] is positive) and contained AO contributions (i.e., both O(2p) and out of plane Pt(5$d_z^2$)) that gave the best $\Delta$Tr[**n(1-n)**] for both calculations simultaneously (Supporting Information Table S16 and Figure S16). This selection of MOPs for the pristine PtO$_2$(110) slab still leads to DFT+U sensitivities that are an order of magnitude higher than the AOP sensitivities for both quantities (Supporting Information Figure S16). Thus, the MOP DFT+U sensitivities are now able to reproduce trends that were previously only achievable with a single parameter when carried out with HF exchange tuning (Figure 9 and Supporting Information Tables S7 and S9).

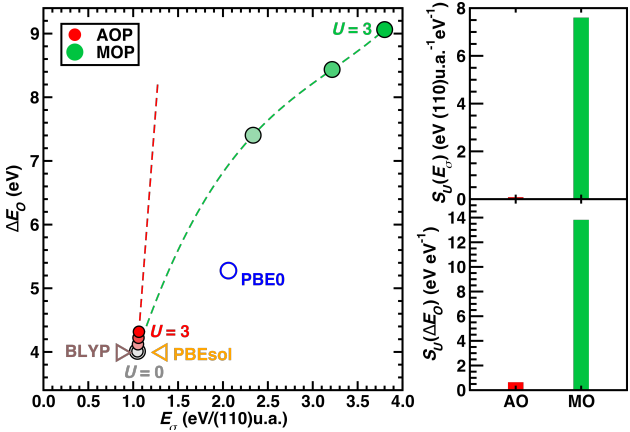

**Figure 9.** (left) $\Delta E_O$ (in eV) versus $E_\sigma$ (in eV/(110)u.a.) for PtO$_2$ (110) computed using the Pt 5$d$ AOPs (shaded red circles) and the fractionality-selected best MOPs (shaded green circles) at integer values of $U$ from 0 to 3 eV with a dashed line indicating the trend between these points. Values from BLYP[129,130] (brown triangle), PBEsol[37] (orange triangle), and PBE0[131,132] (blue circle) were obtained from Ref. [76] using a localized basis set and shifted to align our PBE plane-wave basis set value and the PBE localized basis set value from Ref. [76]. (right) Linearized sensitivities of $E_\sigma$ (top, in eV/(110) unit area per eV of $U$) and of $\Delta E_O$ (bottom, in eV per eV of $U$) computed with AOPs (red bars) and MOPs (green bars).

We used the fractionality-based selection protocol for constructing MOPs for TiO$_2$, a representative early transition-metal oxide where DFT+U with AOPs only tuned $E_\sigma$ (Supporting Information Tables S18–S19). This selection method again yields frontier MOPs from states that lie near the Fermi level (Supporting Information Tables S18–S19). At odds with the standard metal-centered Ti(3*d*) AOPs, the selected MOPs have substantial O(2*p*) contributions (i.e., more than Ti(3*d*)) contributions for both pristine and decorated TiO$_2$(110), especially O(2*p$_z$*) contributions to the MOPs selected for tuning $E_\sigma$ (Supporting Information Tables S18–S19). Similarly, states having majority O*(2*p*) AO character are selected for tuning $\Delta E_O$ of TiO$_2$(110) (Supporting Information Tables S19).

We used these fractionality-selected MOPs to evaluate $E_\sigma$ and $\Delta E_O$ along with their corresponding sensitivities for TiO$_2$ (Figure 10). The MOP-based DFT+U sensitivities are high for both quantities, with magnitudes comparable to HF exchange sensitivities that simultaneously improve both $E_\sigma$ and $\Delta E_O$ with respect to expectations and reference values (Figure 10 and Supporting Information Tables S7 and S9). Similarly, a contiguous "compromise" set of pristine TiO$_2$(110) slab frontier MOPs with majority O(2*p*) AO contributions that maximized the overall $\Delta$Tr[**n(1-n)**] for both adsorption and surface energy calculations enabled simultaneous tuning of both quantities and higher sensitivities than the AOP-based DFT+U approach (Supporting Information Figure S17). Thus, this MOP approach enables an automated strategy to tune both surface formation energies and adsorption energies from their semi-local DFT values regardless of the nature of the transition metal with performance comparable to hybrids but within a low-cost DFT+U framework.

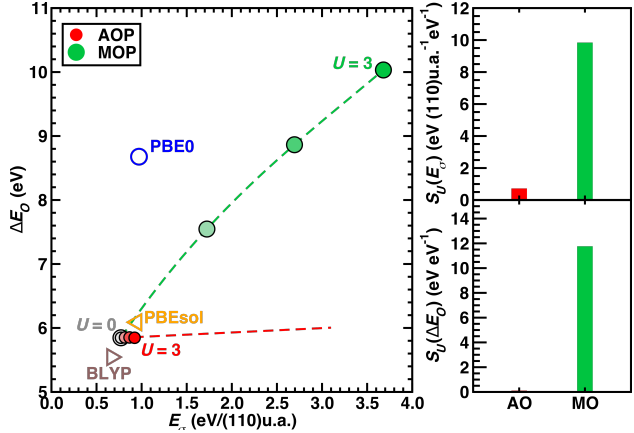

**Figure 10.** (left) $\Delta E_O$ (in eV) versus $E_\sigma$ (in eV/(110)u.a.) for $TiO_2$ (110) computed using the Ti $3d$ AOPs (shaded red circles) and the fractionality-selected best MOPs (shaded green circles) at integer values of $U$ from 0 to 3 eV with a dashed line indicating the trend between these points. Values from BLYP[129,130] (brown triangle), PBEsol[37] (orange triangle), and PBE0[131,132] (blue circle) were obtained from Ref. [76] using a localized basis set and shifted to align our PBE plane-wave basis set value and the PBE localized basis set value from Ref. [76]. (right) Linearized sensitivities of $E_\sigma$ (top, in eV/(110) unit area per eV of $U$) and of $\Delta E_O$ (bottom, in eV per eV of $U$) computed with AOPs (red bars) and MOPs (green bars).

## 4. Conclusions.

We investigated the previously identified limitations of the standard DFT+U approach in simultaneously tuning adsorption energies and surface energies of transition-metal (TM) oxide surfaces. We demonstrated the benefit of using molecular-orbital-like extended projectors (MOPs) for overcoming these limitations within a DFT+U framework. We justified these improvements by analyzing electron density differences between the materials compared in energetic calculations (i.e., the bulk, pristine slabs, and decorated slabs), confirming that standard DFT+U with AOPs tuned the relevant property only when the corresponding density redistribution (e.g., from bulk to pristine slab) was predominantly metal centered. This helped us to rationalize the high sensitivity for surface formation but low sensitivity for adsorption energies in early TMs (e.g., Ti, V) that exhibited metal-local density redistribution during surface formation but a relatively delocalized density redistribution during O atom adsorption, and vice versa for late TMs (e.g., Pt, Ru).

Given the benefit of moving to multi-atom-centered projectors, we developed a systematic protocol for constructing the MOPs to use in DFT+U. Using guiding principles to select sets of frontier orbitals, we were able to identify chemically relevant extended states for MOPs that maximally tuned $E_\sigma$ of the representative late TM $PtO_2$ and $\Delta E_O$ of the representative early TM $TiO_2$. In both cases, the selected MOPs were distinct from the standard metal-centered $d$ AOPs, with the MOPs instead having majority O(2$p$) AO character. Moreover, our protocol highlighted distinctive characteristics of MOPs required for tuning $E_\sigma$ or $\Delta E_O$, with surface O(2$p_z$) contributions key for $E_\sigma$ and adsorbate O*(2$p$) contributions important for $\Delta E_O$ tuning.

We extended our MOP construction scheme for tuning both properties in 3D slabs of $TiO_2$(110) and $PtO_2$(110). The resulting MOP-based DFT+U sensitivities were one to two orders of magnitude higher than those previously obtained using standard DFT+U. This increased sensitivity reduced the value of U needed to tune surface properties, which is beneficial because high values of U in standard DFT+U are known to introduce challenges for convergence of the self-consistent field and over-elongate bonds. The other benefits of this approach are that it maintains a low-cost relative to hybrids and minimizes the number of adjustable parameters in comparison to multi-site DFT+U. We expect this approach to be broadly applicable to a range of surfaces where hybridization is present that normally cannot be addressed with DFT+U. To reduce empiricism, the DFT+U MOPs could be selected to reproduce hybrid functional results or experimental benchmarks, where available.

ASSOCIATED CONTENT

**Supporting Information Available**. This material is available free of charge via the Internet at http://pubs.acs.org.

Lattice parameters of all bulk rutile transition-metal oxides with +U correction; PBE-level metal-adsorbate bond lengths; overview of approach for constructing and using molecular-orbital-like projectors (i.e., Wannier functions) with DFT+U; influence of geometry optimization on energies at low $U$ values; AOP-based DFT+U or HF surface formation energy and adsorption energy sensitivities for all transition-metal oxides; density difference between bulk rutile and pristine surface for all transition-metal oxides near the (110) plane; illustration of the integration scheme used for quantifying density difference between bulk rutile and pristine (110) surface, shown for a representative transition-metal oxide and integral values for all transition-metal oxides; qualitative density difference between pristine and O*-decorated (110) surfaces for oxides of V, Ru, Rh and Ir; quantitative density difference between pristine and O*-decorated (110) surfaces along the M-O* bond for all transition-metal oxides; diagram of unique metal site indexing in all models of $TiO_2$ and $PtO_2$; total number of eigenstates generated at the PBE-level for all models of $TiO_2$ and $PtO_2$ that were used for constructing and testing all possible MOPs; AO contributions of all eigenstates for 2D $TiO_2$ and $PtO_2$; fractionality-based selection of MOPs for tuning 2D $TiO_2$ adsorption energy and PDOS analysis of its fractionality-selected best MOPs; fractionalities and DFT+U sensitivities using AOPs and best MOPs for tuning 2D $PtO_2$ exfoliation energy and 2D $TiO_2$ adsorption energy; PDOS analysis and AO contributions of best MOPs for 3D $PtO_2$ models; AO contributions of best MOPs for 3D $TiO_2$ models; 3D $TiO_2$ and $PtO_2$ surface energetics using consistent MOPs for the pristine model; grid resolutions used for generating all electron density cube files and pseudopotentials used for all calculations (PDF)

Structures of all materials studied in this work (ZIP)

AUTHOR INFORMATION

**Corresponding Author**

*email: hjkulik@mit.edu phone: 617-253-4584

**Notes**

The authors declare no competing financial interest.

ACKNOWLEDGMENT

This material is based upon work supported by the Department of Energy, National Nuclear Security Administration under Award Number DE-NA0003965. This work made use of Department of Defense HPCMP computing resources. H.J.K. holds a Career Award at the Scientific Interface from the Burroughs Wellcome Fund, an AAAS Marion Milligan Mason

Award, and an Alfred P. Sloan award in Chemistry, which supported this work. The authors thank Chenru Duan, Aditya Nandy, Adam H. Steeves, and Vyshnavi Vennelakanti for providing a critical reading of the manuscript.

**For Table of Contents Use Only**

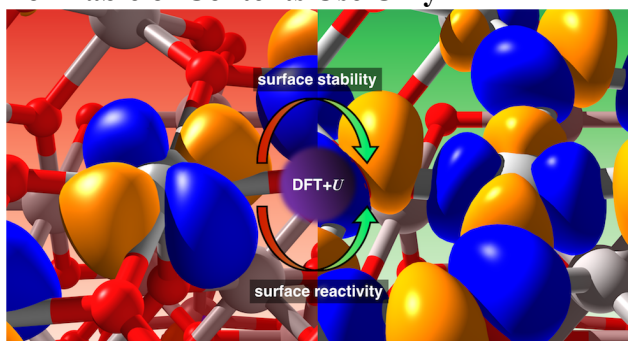

Supporting Information for

*Eliminating Delocalization Error to Improve Heterogeneous Catalysis Predictions with Molecular DFT+U*

Akash Bajaj[1,2] and Heather J. Kulik[1]

[1]Department of Chemical Engineering, Massachusetts Institute of Technology, Cambridge, MA 02139

[2]Department of Materials Science and Engineering, Massachusetts Institute of Technology, Cambridge, MA 02139

**Contents**

**Table S1.** Experimental and geometry-optimized lattice parameters *a* and *c* (in Å) of bulk rutile transition metal oxides $MO_2$ (M = Ti, V, Ru, Rh, Ir, Pt) with *U* (in eV) for DFT+U.

| M | | expt. | PBE | 1 eV | 2 eV | 3 eV | 4 eV | 5 eV | 6 eV | 7 eV | 8 eV | 9 eV | 10 eV |
|---|---|---|---|---|---|---|---|---|---|---|---|---|---|
| Ti | *a* | 4.594 | 4.619 | 4.622 | 4.625 | 4.628 | 4.631 | 4.635 | 4.640 | 4.647 | 4.655 | 4.663 | 4.674 |
|    | *c* | 2.959 | 2.956 | 2.970 | 2.986 | 3.001 | 3.019 | 3.037 | 3.056 | 3.077 | 3.097 | 3.118 | 3.138 |
| V  | *a* | 4.555 | 4.612 | 4.615 | 4.620 | 4.632 | 4.651 | 4.663 | 4.674 | 4.687 | 4.701 | 4.720 | 4.741 |
|    | *c* | 2.853 | 2.778 | 2.779 | 2.777 | 2.766 | 2.755 | 2.765 | 2.770 | 2.777 | 2.787 | 2.799 | 2.817 |
| Ru | *a* | 4.492 | 4.649 | 4.641 | 4.634 | 4.629 | 4.625 | 4.623 | 4.621 | 4.620 | 4.619 | 4.621 | 4.621 |
|    | *c* | 3.107 | 3.190 | 3.191 | 3.193 | 3.193 | 3.194 | 3.192 | 3.191 | 3.190 | 3.190 | 3.188 | 3.188 |
| Rh | *a* | 4.489 | 4.632 | 4.631 | 4.631 | 4.632 | 4.633 | 4.634 | 4.636 | 4.638 | 4.639 | 4.641 | 4.642 |
|    | *c* | 3.090 | 3.174 | 3.172 | 3.170 | 3.168 | 3.167 | 3.166 | 3.165 | 3.164 | 3.163 | 3.163 | 3.163 |
| Ir | *a* | 4.505 | 4.590 | 4.589 | 4.588 | 4.588 | 4.587 | 4.587 | 4.588 | 4.588 | 4.589 | 4.589 | 4.590 |
|    | *c* | 3.159 | 3.214 | 3.211 | 3.208 | 3.205 | 3.202 | 3.200 | 3.197 | 3.195 | 3.193 | 3.191 | 3.189 |
| Pt | *a* | 4.485 | 4.642 | 4.642 | 4.641 | 4.640 | 4.639 | 4.639 | 4.639 | 4.638 | 4.638 | 4.637 | 4.637 |
|    | *c* | 3.130 | 3.280 | 3.275 | 3.271 | 3.268 | 3.265 | 3.264 | 3.262 | 3.262 | 3.262 | 3.263 | 3.264 |

**Table S2.** Geometry-optimized bond lengths between metal adsorption site and O-atom adsorbate (in Å) at the DFT level for the (110) surface of rutile transition metal oxides $MO_2$ (M = Ti, V, Ru, Rh, Ir, Pt). Initial bond lengths were specified using the DFT-level optimized values obtained from Ref. [1] to provide a better initial guess.

| M | Initial | PBE |
|---|---|---|
| Ti | 1.676 | 1.672 |
| V | 1.589 | 1.589 |
| Ru | 1.809 | 1.809 |
| Rh | 1.825 | 1.823 |
| Ir | 1.814 | 1.813 |
| Pt | 1.862 | 1.859 |
| 2D Ti | 1.676 | 1.637 |

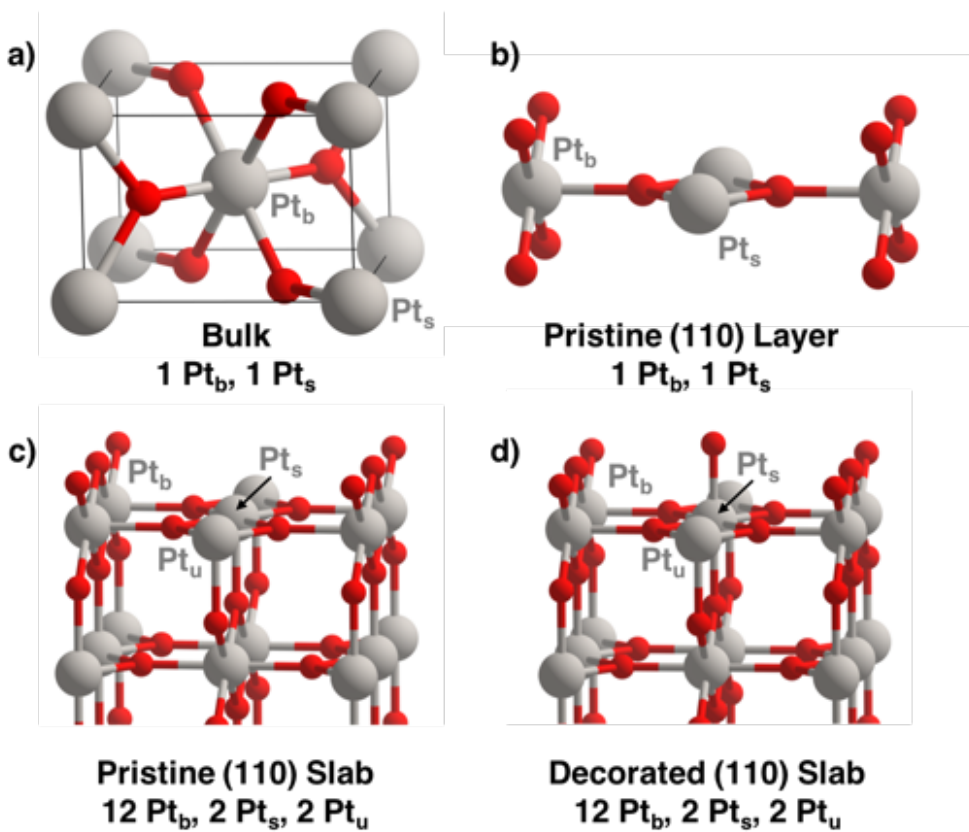

**Figure S1.** Structures of the a) bulk rutile, b) pristine 2D, c) pristine slab, and d) decorated slab models of rutile $PtO_2$. Pt atoms are indicated using gray spheres and oxygen atoms are indicated using red spheres. Pt sites are identified using two or three different types for all DFT+U calculations, where '$Pt_b$' indicates a bulk-like fully coordinated site, '$Pt_u$' indicates a surface undercoordinated site and '$Pt_s$' indicates the site of adsorption. The number of each site within a unit cell is annotated. Note that $Pt_s$ and $Pt_u$ are otherwise equivalent for all pristine surface models.

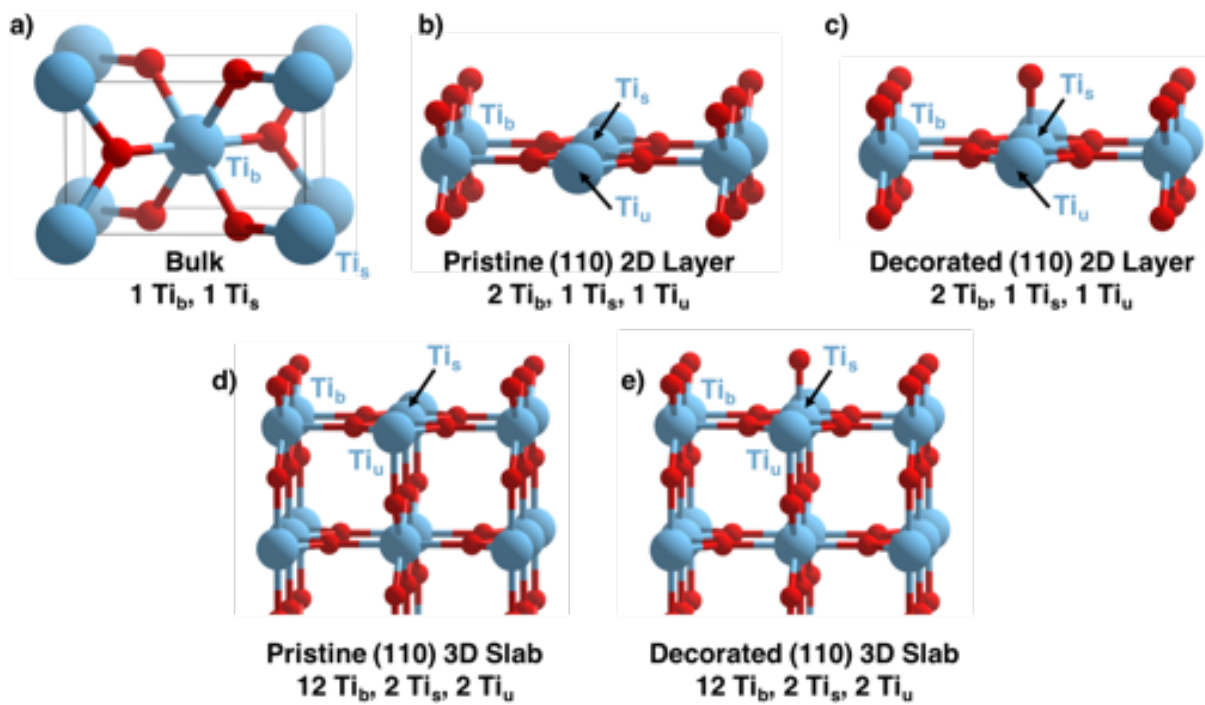

**Figure S2.** Structures of the a) bulk rutile, b) pristine 2D, c) decorated 2D, d) pristine slab and d) decorated slab models of rutile $TiO_2$. Ti atoms are indicated using blue spheres and oxygen atoms are indicated using red spheres. Ti sites are identified using three different types for all DFT+U calculations, where '$Ti_b$' indicates a bulk-like fully coordinated site, '$Ti_u$' indicates a surface undercoordinated site and '$Ti_s$' indicates the site of adsorption. The number of each site within a unit cell is annotated. Note that $Ti_s$ and $Ti_u$ are otherwise equivalent for all pristine surface models.

**Table S3.** List of all pseudopotentials employed in this work with filenames identical to those obtained from the Quantum-ESPRESSO website[2]. The valence electrons that are explicitly modeled are indicated with any semi-core states underlined where applicable.

| Element | Pseudopotential Name | Valence States |
|---|---|---|
| Ti | Ti.pbe-sp-van_ak.UPF | <u>3s</u>, <u>3p</u>, 3d, 4s |
| V | V.pbe-sp-van.UPF | <u>3s</u>, <u>3p</u>, 3d, 4s, 4p |
| Ru | Ru.pbe-n-van.UPF | 4d, 5s, 5p |
| Rh | Rh.pbe-rrkjus.UPF | 4d, 5s |
| Ir | Ir.pbe-n-rrkjus.UPF | 5d, 6s, 6p |
| Pt | Pt.pbe-n-van.UPF | 5d, 6s, 6p |
| O | O.pbe-van_ak.UPF | 2s, 2p |

**Text S1.** Construction of Wannier functions and their use as projectors within DFT+U

The transformation of plane-wave eigenstates $|\psi_{\mathbf{k},v}\rangle$, where $\mathbf{k}$ denotes the *k*-point and *v* denotes the band index, to Wannier functions was carried out using the **pmw.x** utility available with the Quantum-ESPRESSO package[3]. From the total number of available eigenstates, we select a specific set by providing a contiguous range for the band index i.e., $v_i$ to $v_f$. The procedure followed by **pmw.x** after providing it with this selection is as follows:

The selected plane-wave states $\{|\psi_{\mathbf{k},v}\rangle\}$ are first projected on the atomic orbitals (AOs) $\{|\phi_{\lambda'}^I\rangle\}$ localized on a Hubbard site *I*:

$$|\tilde{\psi}_{\mathbf{k},\lambda'}^I\rangle = \sum_{v=v_i}^{v_f} |\psi_{\mathbf{k},v}\rangle \langle \psi_{\mathbf{k},v}|\phi_{\lambda'}^I\rangle, \tag{1}$$

to obtain a set of transformed states $\{|\tilde{\psi}_{\mathbf{k},\lambda'}^I\rangle\}$ corresponding to each Hubbard site *I*. For our transition metal oxides, the AOs were the five valence *d* atomic orbitals of the Hubbard metal site(s), obtained from the all-electron calculation carried out for pseudopotential generation.

The transformed states are then orthonormalized using the Löwdin symmetric orthonormalization procedure:

$$|\bar{\psi}_{\mathbf{k},\lambda}^I\rangle = \sum_{\lambda'} |\tilde{\psi}_{\mathbf{k},\lambda'}^I\rangle \left(S_{\mathbf{k}}^{-\frac{1}{2}}\right)_{\lambda'\lambda}, \tag{2}$$

where $S_{\mathbf{k}}$, the overlap matrix, is defined as $(S_{\mathbf{k}})_{\lambda'\lambda} = \langle \tilde{\psi}_{\mathbf{k},\lambda'}^I | \tilde{\psi}_{\mathbf{k},\lambda}^I \rangle$. The normalized states $\{|\bar{\psi}_{\mathbf{k},\lambda}^I\rangle\}$ are then Fourier transformed from their $\mathbf{k}$-space representation to their real-space representation to obtain the desired Wannier functions for the corresponding Hubbard site *I*:

$$|\mathbf{R}_\lambda^I\rangle = \frac{V}{(2\pi)^3} \int_{BZ} d\mathbf{k}\, e^{-i\mathbf{k}\cdot\mathbf{R}} |\bar{\psi}_{\mathbf{k},\lambda}^I\rangle, \tag{3}$$

where *V* represents the volume of the primitive unit cell in real space.

The generated Wannier functions are used as projectors using the U_projection_type = 'file' command within a DFT+U calculation in Quantum-ESPRESSO, whereupon they replace the AO projectors for only the aforementioned Hubbard site(s) *I*.

**Table S4.** Total number of eigenstates $|\psi_{k,v}\rangle$ generated per **k**-point at the PBE level as specified using the nbnd keyword within Quantum-ESPRESSO for different models of $TiO_2$ and $PtO_2$. "Decorated" refers to a surface model with 0.5 monolayer coverage of oxygen adatoms whereas "pristine" refers to the surface model without the oxygen adatoms.

| Structure | Geometry Model | Total Number of States |
|---|---|---|
| $TiO_2$ | Bulk Rutile | 50 |
| | Pristine (110) 3D Slab | 250 |
| | Pristine (110) 2D Layer | 75 |
| | Decorated (110) 3D Slab | 250 |
| | Decorated (110) 2D Layer | 75 |
| $PtO_2$ | Bulk Rutile | 50 |
| | Pristine (110) 3D Slab | 250 |
| | Pristine (110) 2D Layer | 50 |
| | Decorated (110) 3D Slab | 250 |

**Table S5.** Comparison of DFT+U surface formation energies, $E_\sigma$ (eV/(110) u.a.), for the (110) surface of rutile $TiO_2$ evaluated using optimized geometries at each $U$ and using the DFT-level (i.e. $U = 0$ eV) geometry with single-point energy calculations at each $U$. All calculations employed a Hubbard $U$ on all Ti sites and were carried out using Ti(3d) atomic orbitals as the projector basis within DFT+U.

| $U$ value (eV) | $E_\sigma$ from geometry optimizations | $E_\sigma$ from single-point energies using DFT-level geometry |
|---|---|---|
| 0 | 0.77 | 0.77 |
| 1 | 0.82 | 0.81 |
| 2 | 0.87 | 0.85 |
| 3 | 0.92 | 0.90 |
| 4 | 0.98 | 0.95 |
| 5 | 1.05 | 1.00 |

**Table S6.** Grid resolutions, i.e., the number of grid points along each axis, used to generate electron density cube files of bulk and slab models of rutile transition metal oxides, $MO_2$.

| M | Geometry Model(s) | x Grid Points | y Grid Points | z Grid Points |
|---|---|---|---|---|
| Ti | Bulk Rutile | 200 | 200 | 133 |
| | 3D Slabs | 200 | 266 | 321 |
| V | Bulk Rutile | 200 | 200 | 200 |
| | 3D Slabs | 200 | 400 | 311 |
| Ru | Bulk Rutile | 200 | 200 | 200 |
| | 3D Slabs | 200 | 400 | 328 |
| Rh | Bulk Rutile | 200 | 200 | 200 |
| | 3D Slabs | 200 | 400 | 312 |
| Ir | Bulk Rutile | 200 | 200 | 200 |
| | 3D Slabs | 200 | 400 | 327 |
| Pt | Bulk Rutile | 200 | 200 | 148 |
| | 3D Slabs | 200 | 296 | 325 |

**Table S7.** Linearized sensitivity of the surface formation energy, $S(E_\sigma)$, for the (110) planes of rutile transition metal oxides, $MO_2$. The sensitivities are computed with respect to the $U$ value (in eV/(110) u.a. per eV of $U$) using PBE+U in a plane-wave basis set (and multiplied by 10) or obtained from Ref. [1] (in eV/(110) u.a. per HFX) where it was computed with respect to a change in Hartree–Fock (HF) exchange fraction, $a_{HF}$, from 0 to 1 (i.e., 1 HFX) in the PBE0[4-6] global hybrid functional using an atom-centered basis set. The DFT+U sensitivity for $VO_2$ was obtained from Ref. [1].

| M | DFT+U $S(E_\sigma)$ (10×) (eV (110)u.a.$^{-1}$ eV$^{-1}$) | $a_{HF}$ $S(E_\sigma)$ (eV (110)u.a.$^{-1}$ HFX$^{-1}$) |
|---|---|---|
| Ti | 0.66 | 0.95 |
| V | 0.44 | 0.74 |
| Ru | 0.39 | 3.00 |
| Rh | 0.05 | 1.31 |
| Ir | 0.14 | 0.67 |
| Pt | 0.04 | 3.90 |

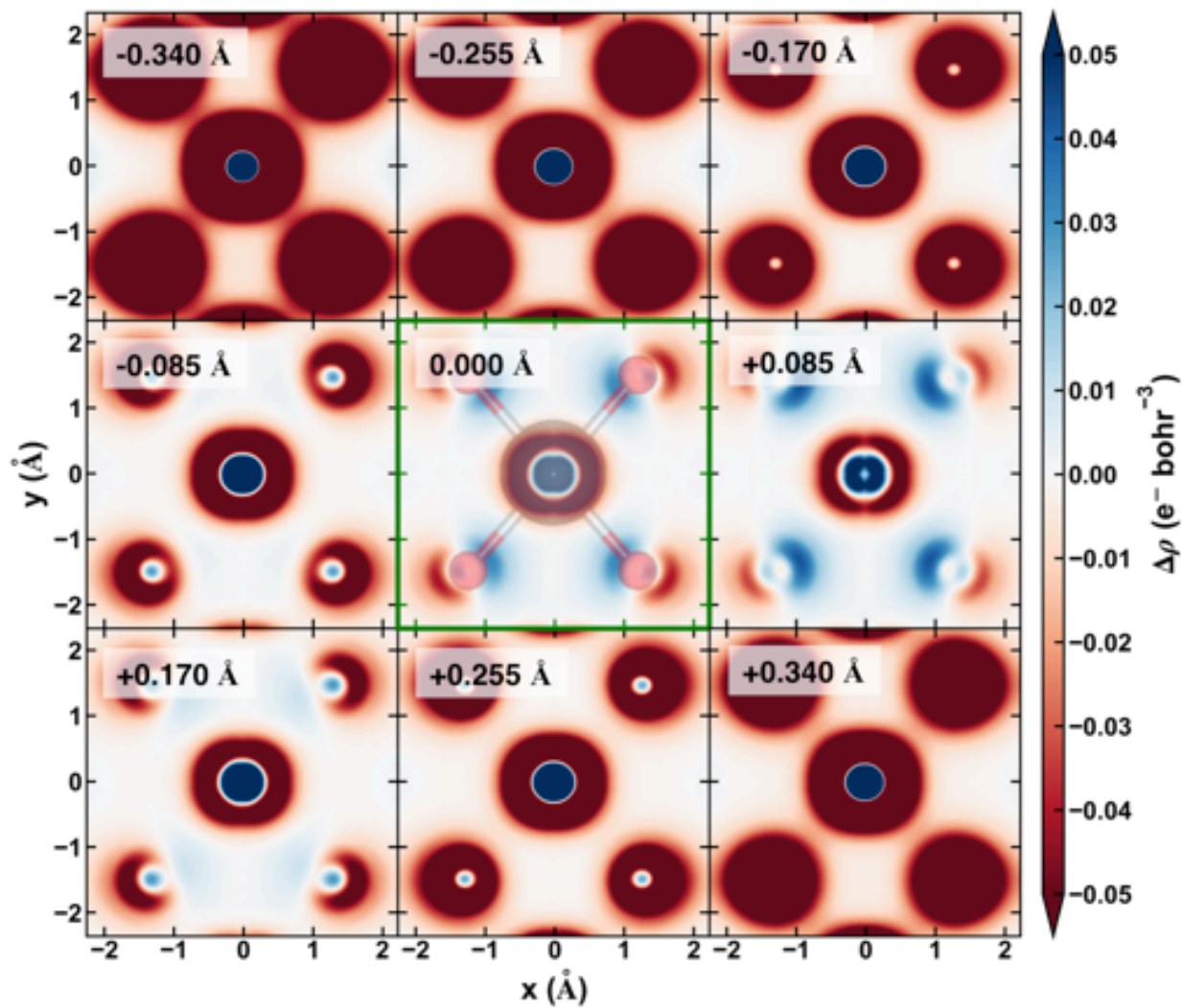

**Figure S3.** Density difference, Δρ, between the pristine slab and the bulk rutile model of $TiO_2$ computed for the (110) plane (middle pane, outlined in green with atom positions shown) and at vertical distances ranging from below the (110) plane (z = 0 Å) at -0.34 Å to above at +0.34 Å as annotated in inset. Red indicates density loss and blue indicates density gain after the surface has been cleaved.

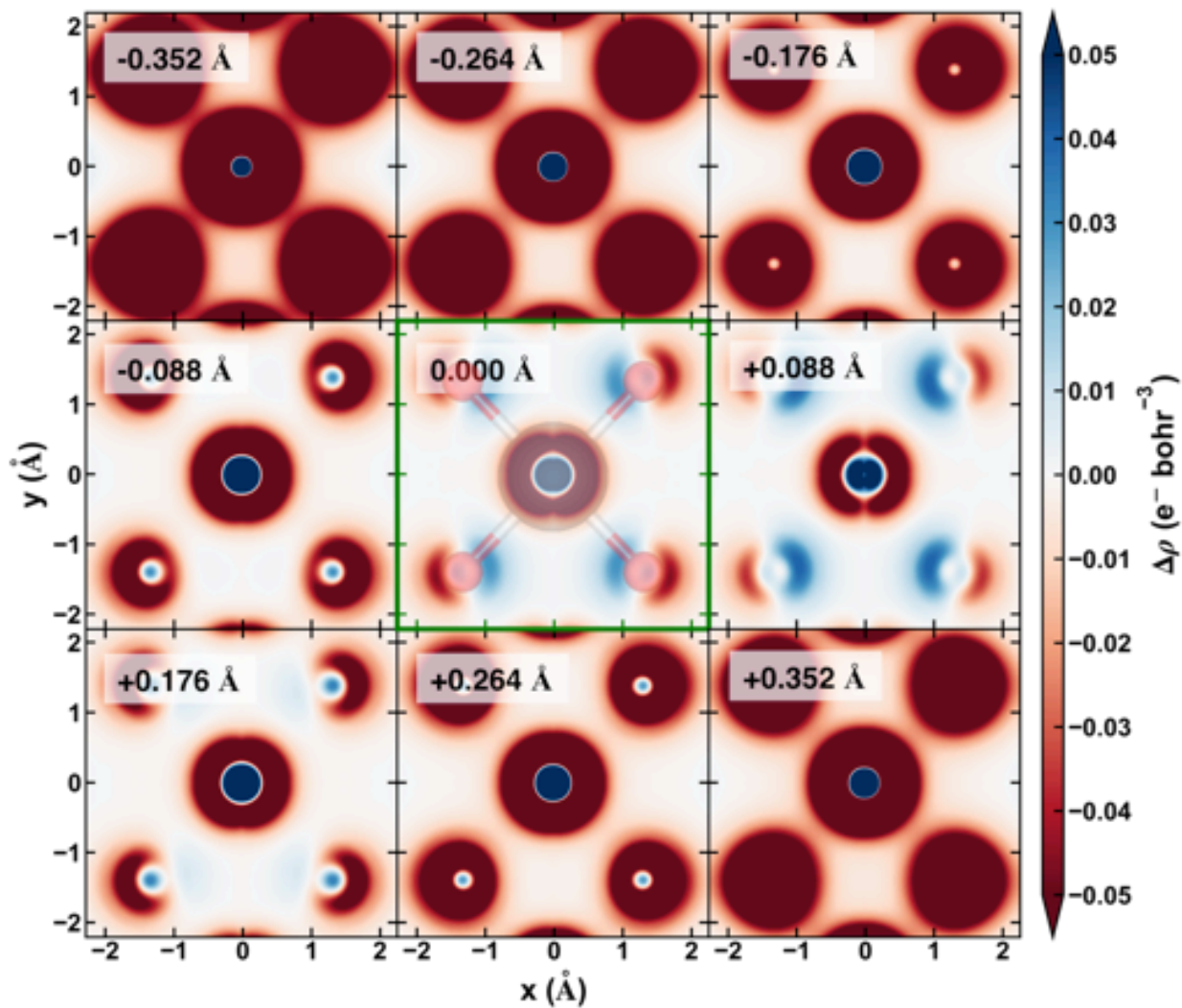

**Figure S4.** Density difference, Δρ, between the pristine slab and the bulk rutile model of VO$_2$ computed for the (110) plane (middle pane, outlined in green with atom positions shown) and at vertical distances ranging from below the (110) plane (z = 0 Å) at -0.352 Å to above at +0.352 Å as annotated in inset. Red indicates density loss and blue indicates density gain after the surface has been cleaved.

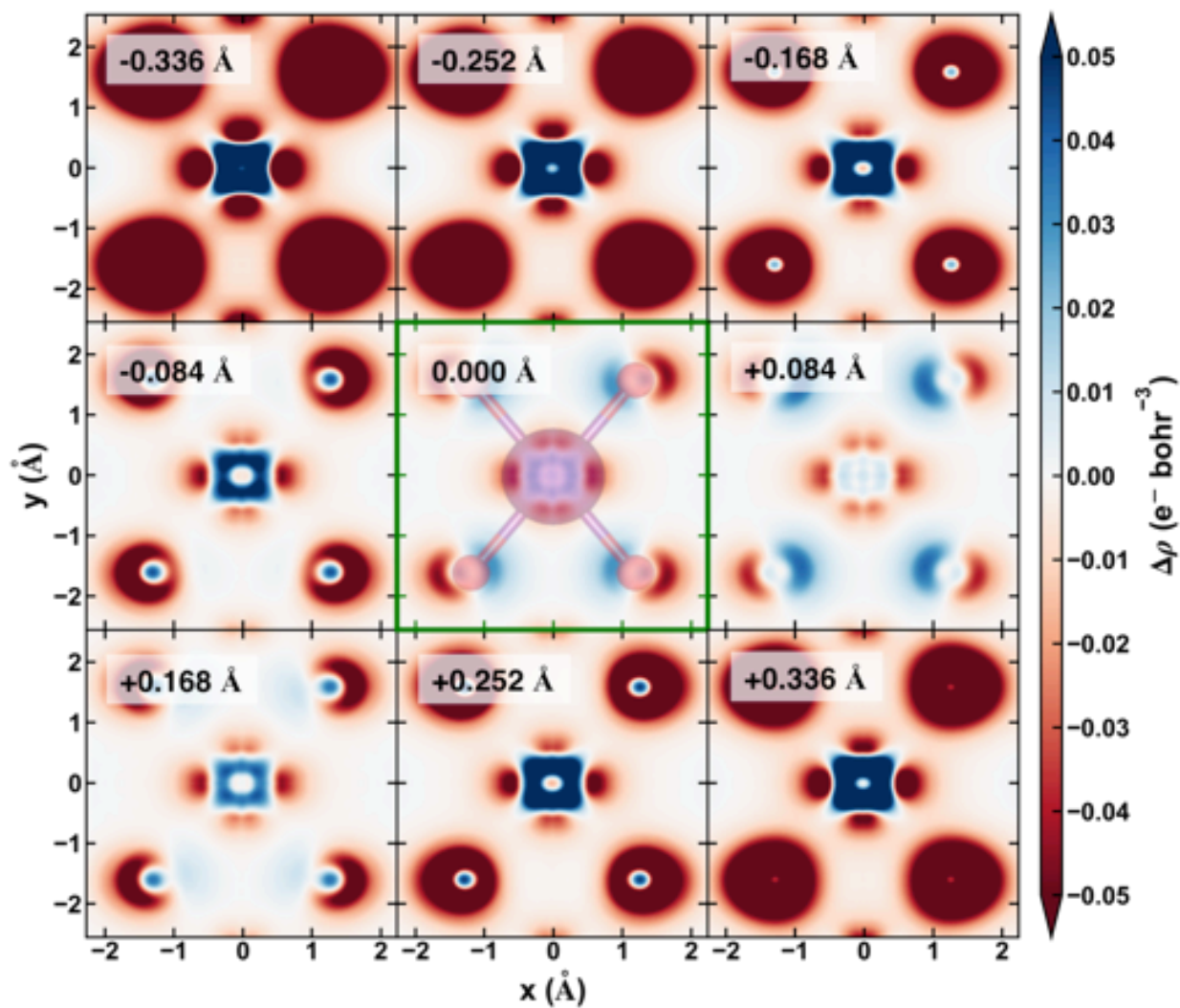

**Figure S5.** Density difference, Δρ, between the pristine slab and the bulk rutile model of RuO$_2$ computed for the (110) plane (middle pane, outlined in green with atom positions shown) and at vertical distances ranging from below the (110) plane (z = 0 Å) at -0.336 Å to above at +0.336 Å as annotated in inset. Red indicates density loss and blue indicates density gain after the surface has been cleaved.

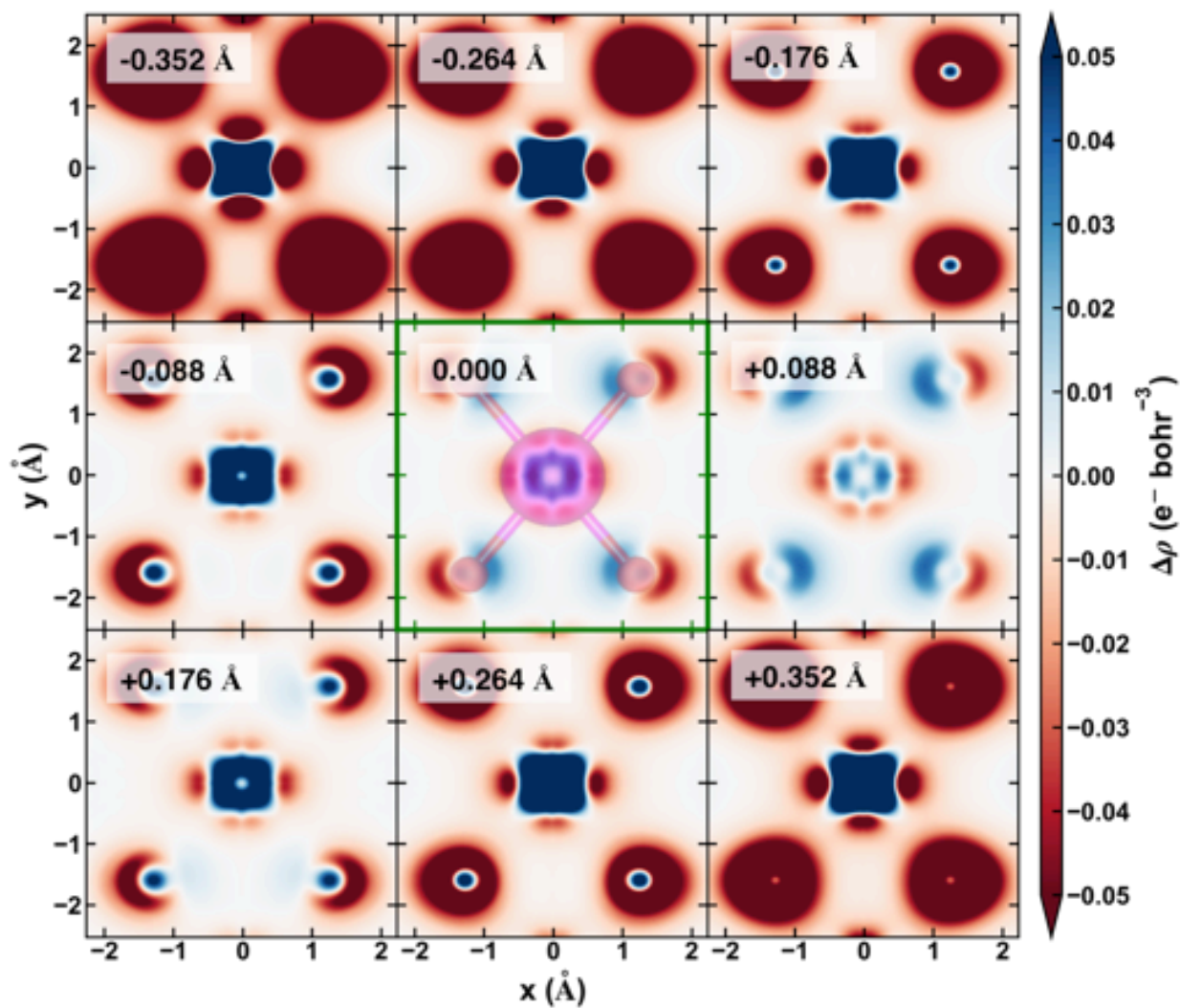

**Figure S6.** Density difference, Δρ, between the pristine slab and the bulk rutile model of RhO$_2$ computed for the (110) plane (middle pane, outlined in green with atom positions shown) and at vertical distances ranging from below the (110) plane (z = 0 Å) at -0.352 Å to above at +0.352 Å as annotated in inset. Red indicates density loss and blue indicates density gain after the surface has been cleaved.

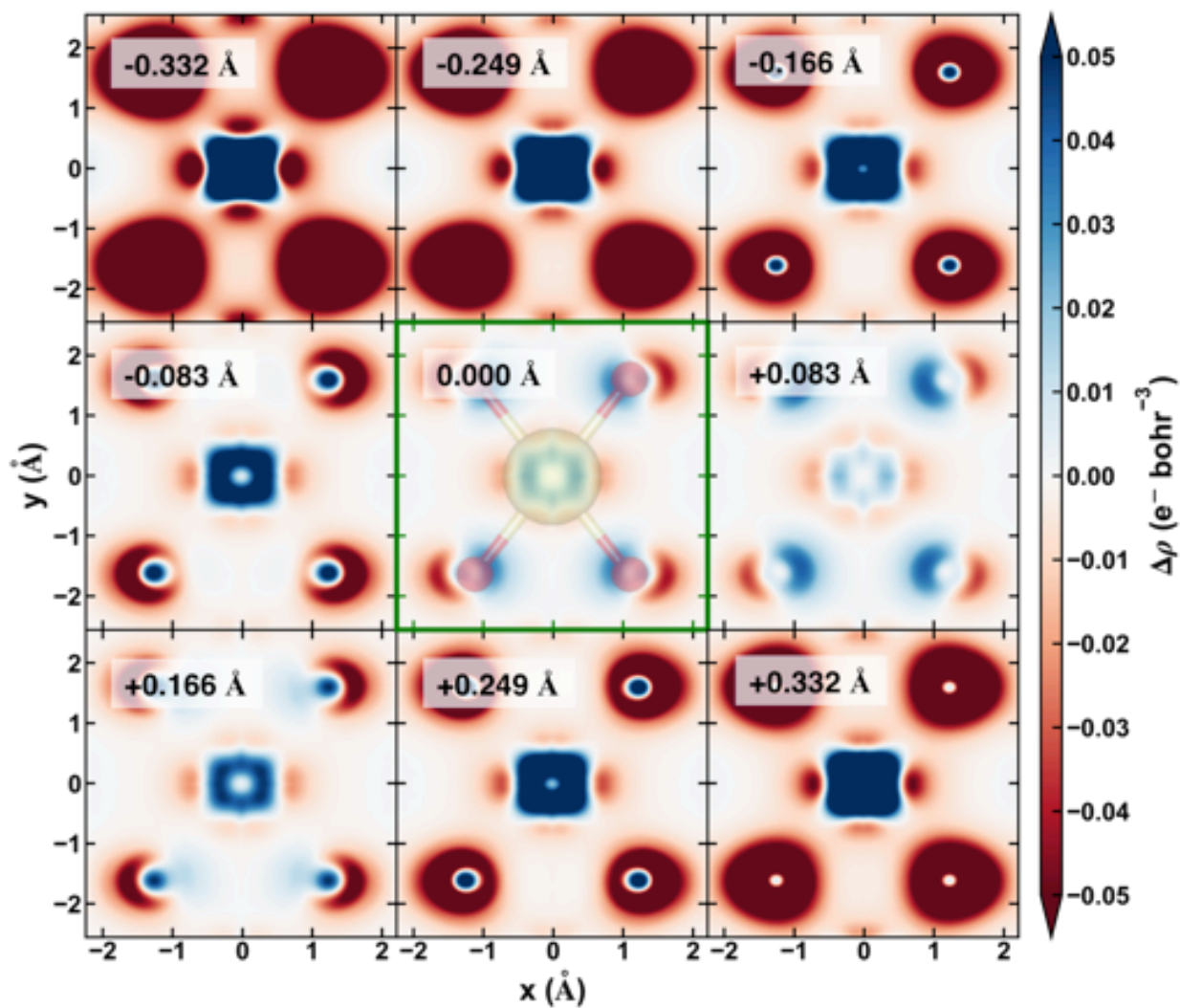

**Figure S7.** Density difference, Δρ, between the pristine slab and the bulk rutile model of IrO$_2$ computed for the (110) plane (middle pane, outlined in green with atom positions shown) and at vertical distances ranging from below the (110) plane (z = 0 Å) at -0.332 Å to above at +0.332 Å as annotated in inset. Red indicates density loss and blue indicates density gain after the surface has been cleaved.

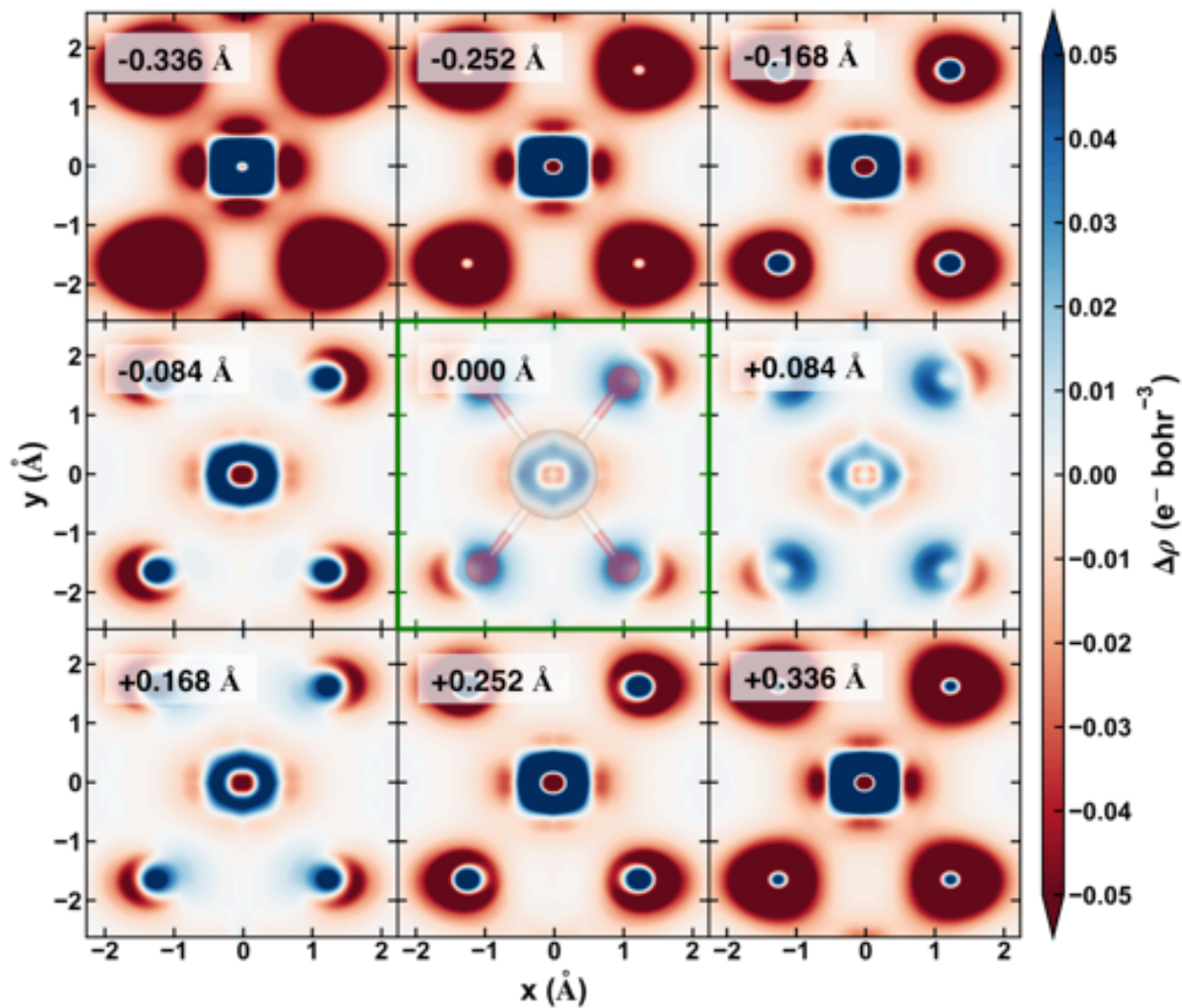

**Figure S8.** Density difference, Δρ, between the pristine slab and the bulk rutile model of $PtO_2$ computed for the (110) plane (middle pane, outlined in green with atom positions shown) and at vertical distances ranging from below the (110) plane (z = 0 Å) at -0.336 Å to above at +0.336 Å as annotated in inset. Red indicates density loss and blue indicates density gain after the surface has been cleaved.

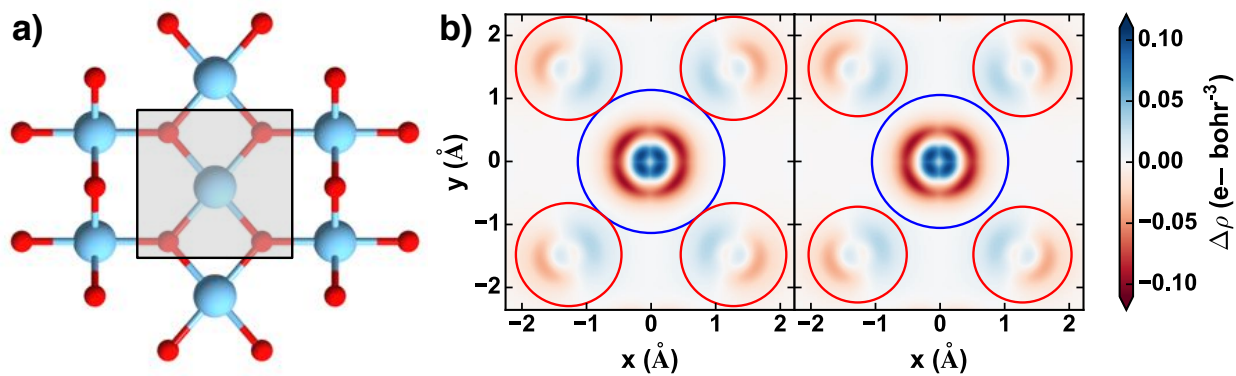

**Figure S9.** (a) Top view of the TiO$_2$(110) surface plane with Ti in blue and O in red. (b) Density difference, $\Delta\rho$, between the pristine slab and the bulk rutile models of TiO$_2$ computed within the shaded region of the (110) plane indicated in (a). The integration of $|\Delta\rho|$ is evaluated within a circular region around the metal (blue circles) and around the oxygen (red circles). The radii of the circles were computed either by scaling down the van der Waals radii[7] of each atom such that the metal and oxygen circles touched (left pane) or by scaling the van der Waals radii[7] of each atom by 0.5x (right pane) to keep a consistent O atom radius for all metals.

**Table S8.** Integral of $|\Delta\rho|$ for MO$_2$(110) surface formation within a circular region centered at the metal site and the sum over all oxygen atoms. All integral values are reported in units of e–/bohr with the fraction of the integral value at the metal site indicated separately. The radii of the circular regions for integration were obtained after scaling the van-der Waals radii[7] of the metal and oxygen atoms, (i) by a system-dependent 'variable scale factor' such that the metal- and oxygen-centered circles touched, and (ii) by a 'fixed scale factor' of 0.5x.

| MO$_2$ | Variable scale factor | | | Fixed scale factor | | |
|---|---|---|---|---|---|---|
| | Metal | Oxygen (total) | Metal contribution | Metal | Oxygen (total) | Metal contribution |
| TiO$_2$ | 0.3790 | 0.3955 | 48.9% | 0.3741 | 0.3661 | 50.5% |
| VO$_2$ | 0.4374 | 0.3837 | 53.3% | 0.4349 | 0.3639 | 54.4% |
| RuO$_2$ | 0.1299 | 0.4021 | 24.4% | 0.1229 | 0.3688 | 25.0% |
| RhO$_2$ | 0.1562 | 0.4067 | 27.8% | 0.1508 | 0.3750 | 28.7% |
| IrO$_2$ | 0.1080 | 0.3946 | 21.5% | 0.1014 | 0.3646 | 21.8% |
| PtO$_2$ | 0.1338 | 0.4358 | 23.5% | 0.1266 | 0.4038 | 23.9% |

**Table S9.** Linearized sensitivity of the adsorption energy, $S(\Delta E_O)$, for adsorption of 0.5 monolayer of O atoms on the (110) planes of rutile transition metal oxides, $MO_2$. The sensitivities are computed with respect to the $U$ value (in eV/eV of $U$) using PBE+U in a plane-wave basis set (x10) or obtained from Ref. [1] (in eV/HFX) where it was computed with respect to a change in Hartree–Fock (HF) exchange fraction, $a_{HF}$, from 0 to 1 (i.e., 1 HFX) in the PBE0[4-6] global hybrid functional using an atom-centered basis set.

| M | DFT+U $S(\Delta E_O)$ (10×) (eV/eV of U) | $a_{HF}$ $S(\Delta E_O)$ (eV/HFX) |
| --- | --- | --- |
| Ti | 0.03 | 11.46 |
| V | 0.00 | 1.96 |
| Ru | 0.49 | 3.21 |
| Rh | 0.64 | 5.88 |
| Ir | 0.91 | 4.38 |
| Pt | 0.56 | 4.58 |

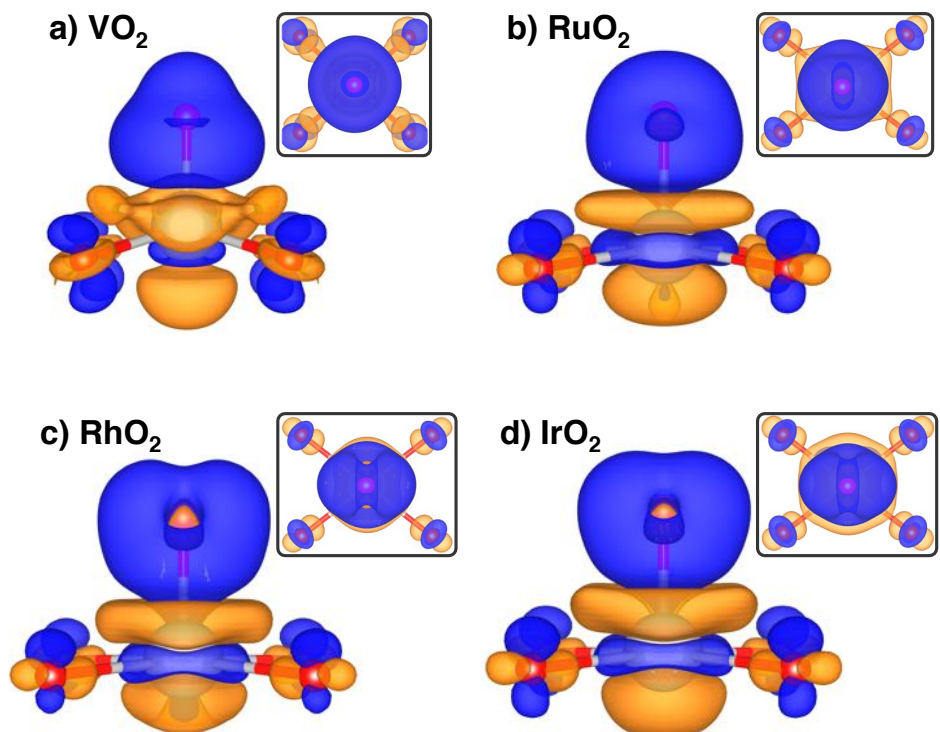

**Figure S10.** Isosurfaces for the density difference (isovalue: 0.003 e⁻/bohr³) between O*-adsorbed and pristine (110) surfaces of (a) $VO_2$, (b) $RuO_2$, (c) $RhO_2$ and (d) $IrO_2$. The metal adsorption site (gray spheres), the four in-plane coordinating O atoms (red spheres) and the O* adsorbate (pink spheres) are indicated in the front view with the top view depicted in the inset. Orange indicates density loss and blue indicates density addition after O* adsorption.

**Table S10.** Maximum absolute value of the density difference i.e., $|\Delta\rho|_{max}$ (in $e^-/Å^3$), between the O*-adsorbed and pristine $MO_2(110)$ surfaces evaluated along the M–O* bond.

| M | $|\Delta\rho|_{max}$ ($e^-/Å^3$) |
|---|---|
| Ti | 0.11 |
| V | 0.19 |
| Ru | 0.05 |
| Rh | 0.07 |
| Ir | 0.07 |
| Pt | 0.06 |

**Table S11.** Fractionality difference per orbital, $\Delta Tr[\mathbf{n}(1\text{-}\mathbf{n})]$ (in $e^{-2}$), between 2D $PtO_2$ (110) and bulk rutile $PtO_2$ computed after summing over both unique Pt sites. $\Delta Tr[\mathbf{n}(1\text{-}\mathbf{n})]$ was computed at the PBE level of theory (i.e., $U = 0$ eV) using atomic projectors at both Pt sites ("Atomic") or after replacing the atomic projectors at the undercoordinated Pt site, $Pt_s$, with fractionality-selected best molecular projectors constructed using the Wannier localization scheme, while retaining atomic projectors at the other Pt site ("Molecular"). Linearized sensitivities of the 2D $PtO_2$ exfoliation energy, $S(E_\sigma)$, computed with respect to the $U$ value (in eV/(110)u.a. per eV of $U$) using PBE+U in a plane-wave basis set (and multiplied by 10) is also tabulated for each projector choice. As expected, the magnitude of $\Delta Tr$ correlates strongly with the magnitude of $S(E_\sigma)$, validating our first-order approximation. However, the sign of the $\Delta Tr$ does not correlate with that of $S(E_\sigma)$ for the atomic projections. This is because the $\Delta Tr$ is too close to zero at $U = 0$ and we noticed a small decrease in $\Delta Tr$ with increasing $U$, which will not be captured within this first-order approximation.

| Projector Type | $\Delta Tr[\mathbf{n}(1\text{-}\mathbf{n})]$ ($e^{-2}$) | DFT+U $S(E_\sigma)$ (10×) (eV (110)u.a.$^{-1}$ eV$^{-1}$) |
|---|---|---|
| Atomic | 0.001 | -0.13 |
| Molecular | 0.088 | 2.96 |

**Table S12.** Atomic orbital (AO) contributions from the Pt$_s$(5d) and O(2p) orbitals for all plane-wave states (i.e., "molecular orbitals", labeled by their band index) available for Wannier function construction in 2D PtO$_2$ (110). All AO contributions are computed at the Γ-point using their projections on the AOs and are listed as a percentage. Pt(5d) and O(2p) columns show the sum of contributions from all the individual d and p AOs respectively. O(2p$_z$) AO contributions are listed separately. The five contiguous states used for constructing molecular projectors within the Wannier localization scheme are indicated in bold.

| Band Index | Pt$_s$(5d) (%) | O(2p) (%) | O(2p$_z$) (%) | Band Index | Pt$_s$(5d) (%) | O(2p) (%) | O(2p$_z$) (%) |
|---|---|---|---|---|---|---|---|
| 1 | 5.1 | 0.6 | 0.2 | 26 | 0 | 60.4 | 0 |
| 2 | 0 | 0 | 0 | 27 | 7.2 | 18.8 | 10.2 |
| 3 | 0.2 | 2.6 | 2.2 | 28 | 0 | 9.6 | 9.6 |
| 4 | 0 | 0.4 | 0.4 | 29 | 0.3 | 0.8 | 0.2 |
| 5 | 19.7 | 54.8 | 0 | 30 | 0 | 1 | 1 |
| 6 | 47.5 | 47.4 | 0 | 31 | 1.2 | 3 | 0 |
| 7 | 0 | 39.4 | 0 | 32 | 2.2 | 16.2 | 4.2 |
| 8 | 0.5 | 40.2 | 40.2 | 33 | 0 | 8.8 | 0 |
| 9 | 12 | 42.8 | 30.4 | 34 | 0 | 0.8 | 0.8 |
| 10 | 0 | 96 | 0 | 35 | 0.2 | 0 | 0 |
| 11 | 0 | 90 | 0 | 36 | 0.2 | 0.6 | 0 |
| 12 | 0 | 89.6 | 89.6 | 37 | 0 | 0.4 | 0 |
| 13 | 7.2 | 0 | 0 | 38 | 0 | 0.4 | 0.4 |
| 14 | 42 | 29.4 | 19.6 | 39 | 0 | 0.4 | 0.4 |
| 15 | 64 | 0.6 | 0 | 40 | 0 | 0 | 0 |
| 16 | 0 | 96.8 | 0 | 41 | 0 | 0.2 | 0 |
| 17 | 99.9 | 0 | 0 | 42 | 0 | 1 | 0.4 |
| 18 | 0 | 94.6 | 0 | 43 | 0.1 | 0 | 0 |
| 19 | 66.7 | 5.2 | 4.6 | 44 | 0 | 0 | 0 |
| **20** | **28.7** | **24.6** | **18** | 45 | 0 | 0 | 0 |
| **21** | **0** | **97** | **97** | 46 | 0 | 0 | 0 |
| **22** | **20.6** | **73.2** | **20.2** | 47 | 0 | 0 | 0 |
| **23** | **24.9** | **54.4** | **29.8** | 48 | 0 | 0.2 | 0.2 |
| **24** | **2.5** | **28.6** | **18.4** | 49 | 0.1 | 0 | 0 |
| 25 | 44.5 | 52 | 0 | 50 | 0 | 4.8 | 0 |

**Table S13.** Fractionality difference per orbital, $\Delta\text{Tr}[\mathbf{n}(1-\mathbf{n})]$ (in $e^{-2}$), between pristine and O*-decorated 2D $TiO_2$(110) computed after summing over all three unique Ti sites. $\Delta\text{Tr}[\mathbf{n}(1-\mathbf{n})]$ was computed at the PBE level of theory (i.e., $U = 0$ eV) using atomic projectors at all three Ti sites ("Atomic"), or after replacing the atomic projectors at the site of adsorption, $Ti_s$, with molecular projectors constructed using the Wannier localization scheme while retaining atomic projectors at the other two Ti sites. Molecular projectors were selected based on the maximum reported $\Delta\text{Tr}$ ("Molecular, $\Delta\text{Tr}_{max}$") or based on physical justification but with a relatively lower $\Delta\text{Tr}$ ("Molecular, Alt."). Linearized sensitivities of the adsorption energy of 0.5 monolayer of O atoms on 2D $TiO_2$(110), $S(\Delta E_O)$, computed with respect to the $U$ value (in eV per eV of $U$) using PBE+U in a plane-wave basis set (and multiplied by 10) is also indicated for each projector choice.

| Projector Type | $\Delta\text{Tr}[\mathbf{n}(1-\mathbf{n})]$ ($e^{-2}$) | DFT+U $S(\Delta E_O)$ (10×) (eV eV$^{-1}$) |
|---|---|---|
| Atomic | 0.006 | -0.06 |
| Molecular, $\Delta\text{Tr}_{max}$ | 0.031 | 5.08 |
| Molecular, Alt. | 0.010 | 4.40 |

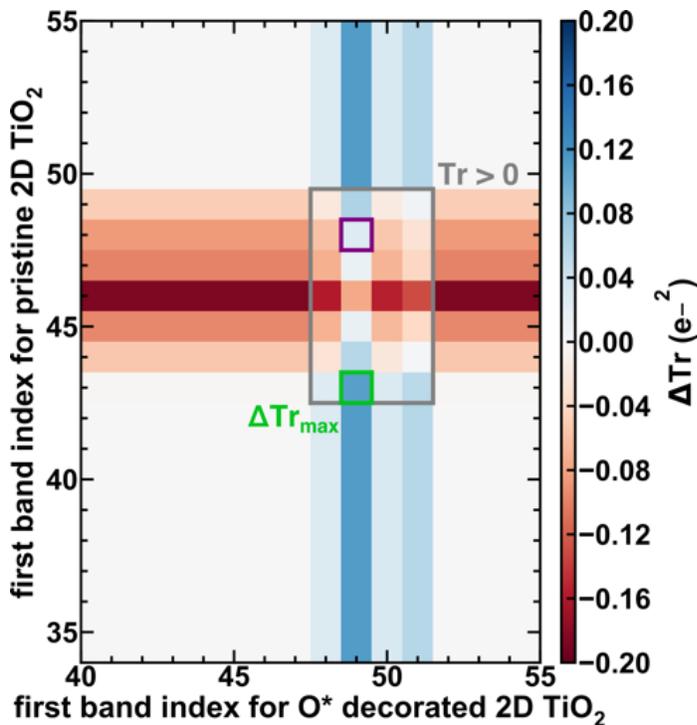

**Figure S11.** Fractionality difference per orbital, $\Delta\text{Tr}[\mathbf{n}(1-\mathbf{n})]$ (in $e^{-2}$), between O*-decorated and pristine 2D $TiO_2$(110) shown for representative molecular projectors constructed using the Wannier localization scheme for both systems. The index of the first band selected for the Wannier localization in O*-decorated 2D $TiO_2$ is shown on the x-axis and the index of the first band selected for the Wannier localization in pristine 2D $TiO_2$ is shown on the y-axis. The projector pair having the maximum $\Delta\text{Tr}[\mathbf{n}(1-\mathbf{n})]$ (green square) from those having a positive Tr (gray rectangle) is highlighted along with an alternative projector pair selected based on physical justification but having relatively lower $\Delta\text{Tr}[\mathbf{n}(1-\mathbf{n})]$ (purple square).

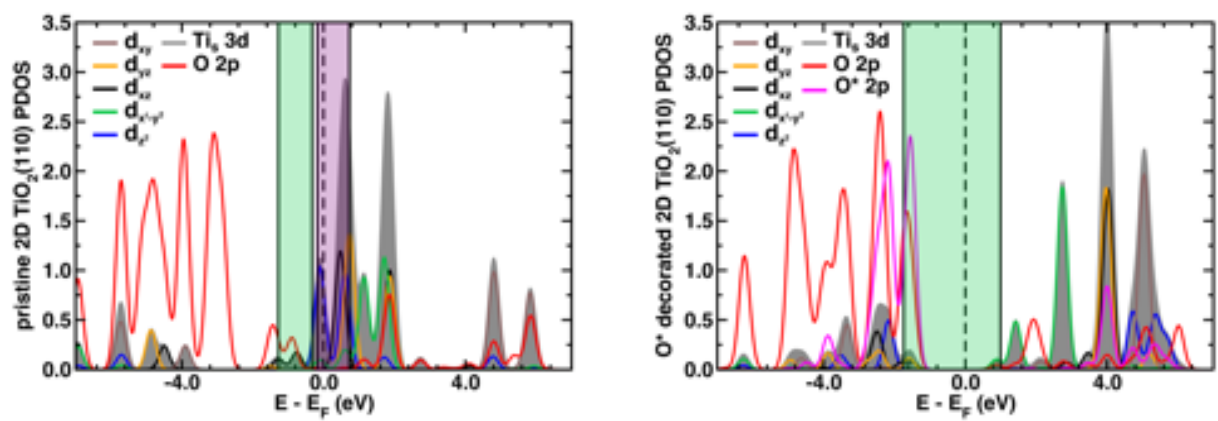

**Figure S12.** Γ-point projected density of states (PDOS) of all Ti$_s$(3$d$) atomic orbitals (grey shaded region) and of all coordinating O(2$p$) atomic orbitals (red solid line) in pristine (left) and O*-decorated (right) 2D TiO$_2$(110). PDOS for the adsorbate O*(2$p$) atomic orbitals is shown in pink and for the Ti$_s$ 3$d_{xy}$, 3$d_{yz}$, 3$d_{xz}$, 3$d_{x^2-y^2}$ and 3$d_{z^2}$ atomic orbitals are shown in brown, orange, black, green and blue respectively. A vertical black dashed line indicates the Fermi level (E$_f$) for each system. Green transparent region represents the fractionality-based best selected states for constructing the molecular projectors within the Wannier localization scheme, whereas the purple transparent region represents the physically motivated alternate selection of states. Note that alternate states were selected for only the pristine 2D TiO$_2$ model.

**Table S14.** Atomic orbital (AO) contributions from the Ti$_s$(3$d$) and O(2$p$) orbitals for all plane-wave states (i.e., "molecular orbitals", labeled by their band index) available for Wannier function construction in pristine 2D TiO$_2$ (110). All AO contributions are computed at the Γ-point using their projections on the AOs and are listed as a percentage. The columns Ti$_s$(3$d$) and O(2$p$) show the sum of contributions from all the individual $d$ and $p$ AOs respectively. Ti$_s$(3$d_{xz}$) and Ti$_s$(3$d_{yz}$) contributions are listed separately. Contributions from O(2$s$) AOs, Ti$_s$(3$s$/3$p$/4$s$) AOs or any Ti$_b$ or Ti$_u$ AOs are not shown. Any other missing contribution was not suitably described by the atomic projections available in the pseudopotentials being used. The five contiguous states selected based on fractionality analysis are highlighted in green and the states selected based on physical justification are highlighted in purple.

| Band Index | Ti$_s$(3$d$) (%) | Ti$_s$(3$d_{xz}$) (%) | Ti$_s$(3$d_{yz}$) (%) | O(2$p$) (%) | Band Index | Ti$_s$(3$d$) (%) | Ti$_s$(3$d_{xz}$) (%) | Ti$_s$(3$d_{yz}$) (%) | O(2$p$) (%) |
|---|---|---|---|---|---|---|---|---|---|
| 1 | 0 | 0 | 0 | 1.2 | 39 | 0 | 0 | 0 | 97.6 |
| 2 | 0 | 0 | 0 | 2 | 40 | 1.1 | 0 | 1.1 | 94.4 |
| 3 | 0 | 0 | 0 | 1.6 | 41 | 0 | 0 | 0 | 83.2 |
| 4 | 0 | 0 | 0 | 2 | 42 | 0.2 | 0 | 0 | 95.2 |
| 5 | 0 | 0 | 0 | 2.4 | 43 | 3.9 | 3.9 | 0 | 75.8 |
| 6 | 0 | 0 | 0 | 0.4 | 44 | 0 | 0 | 0 | 77.6 |
| 7 | 0 | 0 | 0 | 0 | 45 | 0 | 0 | 0 | 98 |
| 8 | 0 | 0 | 0 | 0 | 46 | 6.3 | 6.3 | 0 | 85.8 |
| 9 | 0 | 0 | 0 | 1.6 | 47 | 0 | 0 | 0 | 98 |
| 10 | 0 | 0 | 0 | 2.4 | 48 | 0 | 0 | 0 | 98.4 |
| 11 | 0 | 0 | 0 | 2.4 | 49 | 40.7 | 0 | 0 | 0 |
| 12 | 0 | 0 | 0 | 0.8 | 50 | 43.2 | 43.2 | 0 | 13.2 |
| 13 | 0 | 0 | 0 | 0.8 | 51 | 42.6 | 0 | 0 | 1.6 |
| 14 | 0 | 0 | 0 | 0 | 52 | 49.8 | 0 | 49.8 | 0 |
| 15 | 0 | 0 | 0 | 2 | 53 | 34.1 | 0 | 0 | 7.6 |
| 16 | 0 | 0 | 0 | 1.2 | 54 | 45.3 | 0 | 0 | 6 |
| 17 | 2 | 0 | 0 | 0.4 | 55 | 36.2 | 36.2 | 0 | 25.2 |
| 18 | 5.5 | 0 | 0 | 0 | 56 | 34 | 0 | 34 | 31.2 |
| 19 | 0 | 0 | 0 | 0 | 57 | 4.2 | 0 | 0 | 0.8 |
| 20 | 0 | 0 | 0 | 0 | 58 | 0 | 0 | 0 | 5.6 |
| 21 | 0 | 0 | 0 | 1.2 | 59 | 0.8 | 0 | 0 | 12.4 |
| 22 | 0 | 0 | 0 | 0 | 60 | 0 | 0 | 0 | 2.4 |
| 23 | 0 | 0 | 0 | 0.8 | 61 | 0 | 0 | 0 | 28.4 |
| 24 | 0 | 0 | 0 | 0 | 62 | 1.3 | 1.3 | 0 | 22.4 |
| 25 | 9.8 | 0 | 0 | 66 | 63 | 0 | 0 | 0 | 0.8 |
| 26 | 17.5 | 0 | 0 | 64 | 64 | 5 | 0 | 0 | 3.2 |
| 27 | 7 | 0 | 0 | 75.2 | 65 | 35.6 | 0 | 0 | 19.6 |
| 28 | 0 | 0 | 0 | 85.6 | 66 | 0 | 0 | 0 | 18.4 |
| 29 | 14.3 | 0 | 14.3 | 70.4 | 67 | 0.4 | 0 | 0 | 6 |
| 30 | 0 | 0 | 0 | 84.4 | 68 | 28 | 0 | 0 | 34.8 |
| 31 | 8.6 | 8.6 | 0 | 76 | 69 | 2 | 0 | 0 | 11.2 |
| 32 | 0 | 0 | 0 | 96.4 | 70 | 0 | 0 | 0 | 0 |
| 33 | 8.7 | 0 | 0 | 78.8 | 71 | 0 | 0 | 0 | 16 |
| 34 | 0 | 0 | 0 | 68 | 72 | 0 | 0 | 0 | 5.6 |
| 35 | 0 | 0 | 0 | 73.6 | 73 | 2.5 | 0 | 0 | 6.4 |
| 36 | 0 | 0 | 0 | 96.8 | 74 | 0 | 0 | 0 | 36.4 |
| 37 | 0 | 0 | 0 | 63.6 | 75 | 0 | 0 | 0 | 0.4 |
| 38 | 0 | 0 | 0 | 93.6 | | | | | |

**Table S15.** Atomic orbital (AO) contributions from the Ti$_s$(3$d$), O(2$p$) and O*(2$p$) orbitals for all plane-wave states (i.e., "molecular orbitals", labeled by their band index) available for Wannier function construction in O*-decorated 2D TiO$_2$ (110). All AO contributions are computed at the Γ-point using their projections on the AOs and are given as a percentage. The Ti$_s$(3$d$) and O(2$p$) or O*(2$p$) columns give the sum of contributions from all the individual $d$ and $p$ AOs respectively. Ti$_s$(3$d_{xz}$) and Ti$_s$(3$d_{yz}$) contributions are listed separately. Contributions from O(2$s$) AOs, Ti$_s$(3$s$/3$p$/4$s$) AOs or any Ti$_b$ or Ti$_u$ AOs are not shown. Any other missing contribution was not suitably described by the atomic projections available in the pseudopotentials being used. The five contiguous states selected based on fractionality analysis are highlighted in green.

| Band Index | Ti$_s$(3$d$) (%) | Ti$_s$(3$d_{xz}$) (%) | Ti$_s$(3$d_{yz}$) (%) | O*(2$p$) (%) | O(2$p$) (%) | Band Index | Ti$_s$(3$d$) (%) | Ti$_s$(3$d_{xz}$) (%) | Ti$_s$(3$d_{yz}$) (%) | O*(2$p$) (%) | O(2$p$) (%) |
|---|---|---|---|---|---|---|---|---|---|---|---|
| 1 | 0 | 0 | 0 | 0 | 1.6 | 39 | 2.5 | 0 | 0 | 0.6 | 82.1 |
| 2 | 0 | 0 | 0 | 0 | 1.7 | 40 | 16.2 | 0 | 0 | 0 | 78.8 |
| 3 | 0 | 0 | 0 | 0 | 3 | 41 | 0.4 | 0 | 0 | 0.2 | 77.1 |
| 4 | 0 | 0 | 0 | 1.1 | 0.8 | 42 | 3 | 0 | 3 | 8.7 | 83.3 |
| 5 | 0 | 0 | 0 | 0 | 2.9 | 43 | 4 | 4 | 0 | 2.8 | 80 |
| 6 | 0 | 0 | 0 | 0 | 2 | 44 | 11.1 | 11.1 | 0 | 32 | 50.7 |
| 7 | 0 | 0 | 0 | 0 | 0 | 45 | 6.6 | 0 | 6.6 | 6.7 | 78.9 |
| 8 | 0 | 0 | 0 | 0 | 1.2 | 46 | 0.8 | 0 | 0 | 0 | 89.2 |
| 9 | 0 | 0 | 0 | 0 | 0.8 | 47 | 17.9 | 0 | 0 | 69.6 | 5.2 |
| 10 | 0 | 0 | 0 | 0 | 1.8 | 48 | 2.3 | 2.3 | 0 | 6.4 | 87.2 |
| 11 | 0 | 0 | 0 | 0 | 2.4 | **49** | **0.2** | **0.2** | **0** | **0.1** | **96.2** |
| 12 | 0 | 0 | 0 | 0 | 0.8 | **50** | **4.2** | **0** | **4.2** | **52.8** | **40.2** |
| 13 | 0 | 0 | 0 | 0 | 2.6 | **51** | **2.5** | **2.5** | **0** | **35.5** | **54.7** |
| 14 | 0 | 0 | 0 | 2.1 | 0 | **52** | **2.9** | **0** | **0** | **0.1** | **13.1** |
| 15 | 0 | 0 | 0 | 0.2 | 0.8 | **53** | **1** | **0** | **0** | **0** | **84.1** |
| 16 | 0 | 0 | 0 | 0.2 | 0.8 | 54 | 17.7 | 0 | 0 | 0 | 4 |
| 17 | 0 | 0 | 0 | 0 | 7.6 | 55 | 0.3 | 0.3 | 0 | 0.5 | 20.7 |
| 18 | 0.7 | 0 | 0 | 0 | 0 | 56 | 0.6 | 0 | 0.6 | 0 | 20 |
| 19 | 2.8 | 0 | 0 | 0 | 0 | 57 | 0.3 | 0 | 0 | 0 | 3.6 |
| 20 | 0 | 0 | 0 | 0 | 0.8 | 58 | 0.8 | 0 | 0 | 0 | 15.9 |
| 21 | 0.3 | 0 | 0.3 | 0 | 0 | 59 | 2.7 | 0 | 0 | 0 | 6 |
| 22 | 0 | 0 | 0 | 0 | 0.4 | 60 | 66.5 | 0 | 0 | 0 | 4 |
| 23 | 0.2 | 0.2 | 0 | 0 | 0.6 | 61 | 2.6 | 2.6 | 0 | 0.7 | 25.3 |
| 24 | 0 | 0 | 0 | 0 | 3 | 62 | 6 | 6 | 0 | 2.4 | 21 |
| 25 | 5.4 | 0 | 0 | 0.2 | 0 | 63 | 34.4 | 0 | 34.4 | 7.6 | 11.4 |
| 26 | 0.1 | 0 | 0 | 0 | 80.8 | 64 | 34.4 | 0 | 34.4 | 8.4 | 13.2 |
| 27 | 5.3 | 0 | 0 | 0 | 66.6 | 65 | 65.9 | 65.9 | 0 | 15.5 | 7.3 |
| 28 | 0.1 | 0.1 | 0 | 0 | 80.7 | 66 | 0.8 | 0 | 0 | 0.7 | 3.9 |
| 29 | 0 | 0 | 0 | 0 | 82.2 | 67 | 21.9 | 0 | 0 | 7.4 | 9.6 |
| 30 | 0.7 | 0 | 0.7 | 0.2 | 68.9 | 68 | 71.2 | 0 | 0 | 0 | 20 |
| 31 | 2.9 | 0 | 2.9 | 0.4 | 74.8 | 69 | 3.4 | 0 | 0 | 1 | 0.8 |
| 32 | 2.7 | 0 | 0 | 0 | 77.7 | 70 | 5.7 | 0 | 5.7 | 1.8 | 20.5 |
| 33 | 4.1 | 0 | 0 | 0 | 66.8 | 71 | 18.1 | 0 | 0 | 7.7 | 3.2 |
| 34 | 2.6 | 2.6 | 0 | 2.6 | 73.3 | 72 | 11.3 | 0 | 0 | 3.6 | 8.4 |
| 35 | 0.8 | 0.8 | 0 | 0.1 | 91.7 | 73 | 1.1 | 0 | 0 | 0.3 | 20.5 |
| 36 | 5.9 | 0 | 5.9 | 12.1 | 68.5 | 74 | 1.4 | 0 | 0 | 0 | 31.2 |
| 37 | 0 | 0 | 0 | 0.2 | 75.9 | 75 | 0 | 0 | 0 | 0 | 31.3 |
| 38 | 3 | 0 | 0 | 1.2 | 83.7 | | | | | | |

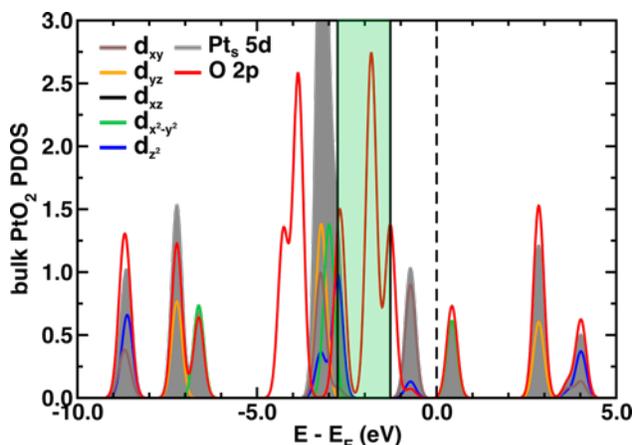

**Figure S13.** Γ-point projected density of states (PDOS) of all Pt$_s$(5$d$) atomic orbitals (gray shaded region) and of the coordinating O(2$p$) atomic orbitals in the (110) plane (red solid line) for bulk rutile PtO$_2$. The PDOS for the Pt$_s$ 5$d_{xy}$, 5$d_{yz}$, 5$d_{xz}$, 5$d_{x^2-y^2}$, and 5$d_{z^2}$ atomic orbitals are shown in brown, orange, black, green, and blue, respectively. The vertical black dashed line indicates the Fermi level (E$_f$). The green transparent region represents the fractionality-based best selected states for constructing the molecular projectors within the Wannier localization scheme for improving surface energy sensitivity.

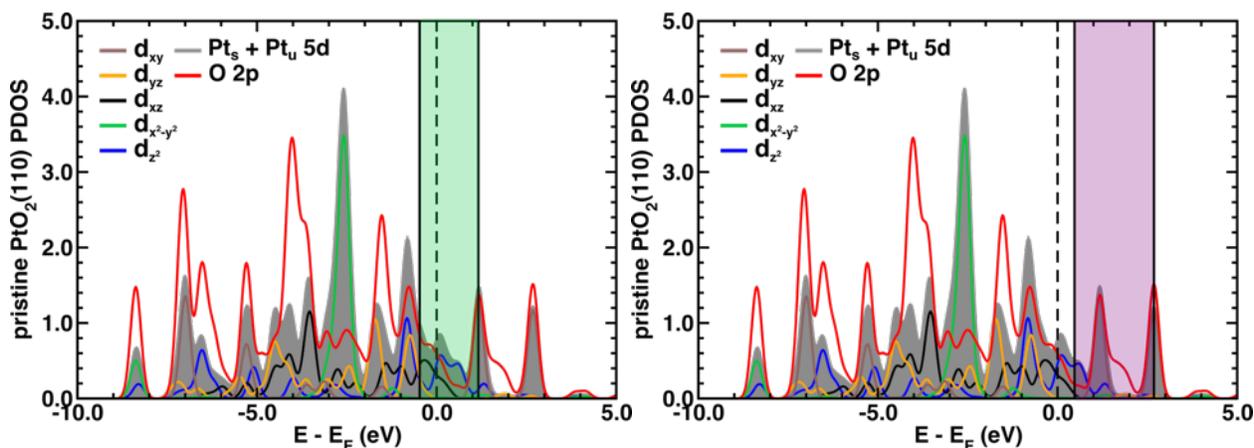

**Figure S14.** Γ-point projected density of states (PDOS) of all Pt$_s$(5$d$) and Pt$_u$(5$d$) atomic orbitals (grey shaded region) and of the coordinating O(2$p$) atomic orbitals in the (110) plane (red solid line) for pristine PtO$_2$(110). The PDOS for the Pt 5$d_{xy}$, 5$d_{yz}$, 5$d_{xz}$, 5$d_{x^2-y^2}$, and 5$d_{z^2}$ atomic orbitals are shown in brown, orange, black, green, and blue, respectively. The vertical black dashed line indicates the Fermi level (E$_f$). The green transparent region (left) and purple transparent region (right) represents the fractionality-based best selected states for constructing the molecular projectors within the Wannier localization scheme for improving surface energy sensitivity and for improving adsorption energy sensitivity, respectively.

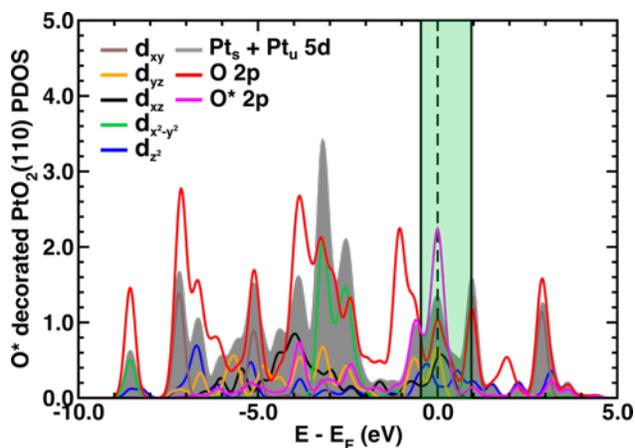

**Figure S15.** Γ-point projected density of states (PDOS) of all $Pt_s(5d)$ and $Pt_u(5d)$ atomic orbitals (grey shaded region) and of all coordinating $O(2p)$ atomic orbitals in the (110) plane (red solid line) in O*-decorated $PtO_2$(110). The PDOS for the adsorbate $O^*(2p)$ atomic orbitals are shown in pink and for the Pt $5d_{xy}$, $5d_{yz}$, $5d_{xz}$, $5d_{x^2-y^2}$, and $5d_{z^2}$ atomic orbitals are shown in brown, orange, black, green, and blue, respectively. The vertical black dashed line indicates the Fermi level ($E_f$) for each system. The green transparent region represents the fractionality-based best selected states for constructing the molecular projectors within the Wannier localization scheme for improving adsorption energy sensitivity.

**Table S16.** Atomic orbital (AO) contributions from the $Pt_s(5d)$, $Pt_u(5d)$, and top surface layer $O(2p)$ orbitals for the fractionality-based best selected plane-wave states (i.e., "molecular orbitals", labeled by their band index) for Wannier function construction in pristine $PtO_2(110)$ for improving surface energy sensitivity and for improving adsorption energy sensitivity. All AO contributions are computed at the Γ-point using their projections on the AOs and are given as a percentage. The columns $Pt(5d)$ and $O(2p)$ contain the sum of contributions from all the individual $d$ and $p$ AOs respectively. $O(2p_z)$ AO contributions are listed separately. Contributions from $O(2s)$ AOs, $Pt_s(6s/6p)$ AOs, $Pt_u(6s/6p)$ AOs or from the AOs of any other Pt or O atoms in the surface model are not shown. Any missing contribution (i.e., rows that sum up to less than 100%) was not suitably described by the atomic projections available in the pseudopotentials being used. The Fermi level lies between the bands indexed 174 (i.e., the HOMO) and 175 (i.e., the LUMO), both of which are highlighted in bold.

| Surface Energy Sensitivity | | | | | Adsorption Energy Sensitivity | | | | |
|---|---|---|---|---|---|---|---|---|---|
| Band Index | $Pt_s(5d)$ (%) | $Pt_u(5d)$ (%) | $O(2p)$ (%) | $O(2p_z)$ (%) | Band Index | $Pt_s(5d)$ (%) | $Pt_u(5d)$ (%) | $O(2p)$ (%) | $O(2p_z)$ (%) |
| 168 | 5.9 | 5.9 | 16.8 | 9.6 | 182 | 3 | 3 | 8.2 | 7 |
| 169 | 3.8 | 3.8 | 8 | 8 | 183 | 6.7 | 6.7 | 4.4 | 4.4 |
| 170 | 1.3 | 1.3 | 13.4 | 0 | 184 | 1.5 | 1.5 | 2.2 | 1.4 |
| 171 | 1.3 | 1.3 | 3.8 | 2.8 | 185 | 0 | 0 | 0 | 0 |
| 172 | 4.7 | 4.7 | 10.4 | 10 | 186 | 0 | 0 | 0 | 0 |
| 173 | 1.7 | 1.7 | 10.8 | 7.2 | 187 | 23.9 | 23.9 | 41.2 | 8.8 |
| **174** | **0** | **0** | **9.2** | **2.8** | 188 | 0 | 0 | 0 | 0 |
| **175** | **1.1** | **1.1** | **1.4** | **0** | 189 | 0.5 | 0.5 | 12 | 8 |
| 176 | 8.9 | 8.9 | 6.8 | 6.8 | 190 | 1.5 | 1.5 | 8 | 7.2 |
| 177 | 0.5 | 0.5 | 21.4 | 0 | 191 | 1.7 | 1.7 | 7.6 | 6.8 |
| 178 | 1.9 | 1.9 | 12.2 | 6 | 192 | 1.1 | 1.1 | 13.6 | 6.8 |
| 179 | 0.5 | 0.5 | 15.4 | 0 | 193 | 0 | 0 | 8.2 | 0.8 |
| 180 | 3.2 | 3.2 | 6 | 5.2 | 194 | 0.3 | 0.3 | 14.8 | 2 |
| 181 | 1.2 | 1.2 | 6 | 2.8 | 195 | 0.7 | 0.7 | 10.2 | 1.6 |
| 182 | 3 | 3 | 8.2 | 7 | 196 | 0.4 | 0.4 | 3.6 | 0.4 |
| 183 | 6.7 | 6.7 | 4.4 | 4.4 | 197 | 0.9 | 0.9 | 2 | 0 |
| 184 | 1.5 | 1.5 | 2.2 | 1.4 | 198 | 0.6 | 0.6 | 0 | 0 |
| 185 | 0 | 0 | 0 | 0 | 199 | 0 | 0 | 1.4 | 0 |
| 186 | 0 | 0 | 0 | 0 | 200 | 0.3 | 0.3 | 3 | 3 |
| 187 | 23.9 | 23.9 | 41.2 | 8.8 | 201 | 21.1 | 21 | 54 | 0 |

**Table S17.** Atomic orbital (AO) contributions from the $Pt_s(5d)$, $Pt_u(5d)$, $O^*(2p)$ and top surface layer $O(2p)$ orbitals for the fractionality-based best selected plane-wave states (i.e., "molecular orbitals", labeled by their band index) for Wannier function construction in $O^*$-decorated $PtO_2(110)$ for improving adsorption energy sensitivity. All AO contributions are computed at the $\Gamma$-point using their projections on the AOs and are given as a percentage. The columns $Pt(5d)$ and $O(2p)$ contain the sum of contributions from all the individual $d$ and $p$ AOs respectively. Contributions from $O(2s)$ AOs, $Pt_s(6s/6p)$ AOs, $Pt_u(6s/6p)$ AOs or from the AOs of any other Pt or O atoms in the surface model are not shown. Any other missing contribution (i.e., rows that sum up to less than 100%) was not suitably described by the atomic projections available in the pseudopotentials being used. The Fermi level lies between the bands indexed 177 (i.e., the HOMO) and 178 (i.e., the LUMO), both of which are highlighted in bold.

| Band Index | $Pt_s(5d)$ (%) | $Pt_u(5d)$ (%) | $O(2p)$ (%) | $O^*(2p)$ (%) |
|---|---|---|---|---|
| 170 | 0.3 | 3.9 | 7.8 | 13.8 |
| 171 | 0.4 | 2.5 | 15.7 | 1 |
| 172 | 0.2 | 14.3 | 3.7 | 1.1 |
| 173 | 0.7 | 1.8 | 2.9 | 6.9 |
| 174 | 1.6 | 0 | 0 | 7.4 |
| 175 | 2.4 | 1.5 | 9 | 2 |
| 176 | 3.2 | 0.5 | 9.6 | 12.8 |
| **177** | **18** | **4.9** | **15.8** | **44.8** |
| **178** | **3.8** | **1** | **13.6** | **2.2** |
| 179 | 1.3 | 1.9 | 8.5 | 3.3 |
| 180 | 1.9 | 2.8 | 2.7 | 3.7 |
| 181 | 3.8 | 0.3 | 9.9 | 15.8 |
| 182 | 0.2 | 0.6 | 18.4 | 0.6 |
| 183 | 3.1 | 4.1 | 19.1 | 4.4 |
| 184 | 2.3 | 3 | 9.6 | 2.1 |
| 185 | 0 | 2 | 10.9 | 0.4 |
| 186 | 0.8 | 10.3 | 4.5 | 0.2 |
| 187 | 1.6 | 0.9 | 4.4 | 1.6 |
| 188 | 0.3 | 0.2 | 0 | 0 |
| 189 | 0 | 0 | 0 | 0 |

**Table S18.** Atomic orbital (AO) contributions from the $Ti_s(3d)$, $Ti_u(3d)$, and top surface layer $O(2p)$ orbitals for the fractionality-based best selected plane-wave states (i.e., "molecular orbitals", labeled by their band index) for Wannier function construction in pristine $TiO_2(110)$ for improving surface energy sensitivity and for improving adsorption energy sensitivity. All AO contributions are computed at the Γ-point using their projections on the AOs and are given as a percentage. The $Ti(3d)$ and $O(2p)$ columns contain the sum of contributions from all the individual $d$ and $p$ AOs respectively. $O(2p_z)$ AO contributions are listed separately. Contributions from $O(2s)$ AOs, $Ti_s(3s/3p/4s)$ AOs, $Ti_u(3s/3p/4s)$ AOs or from the AOs of any other Ti or O atoms in the surface model are not shown. Any missing contribution (i.e., rows that sum up to less than 100%) was not suitably described by the atomic projections available in the pseudopotentials being used. The Fermi level lies between the bands indexed 192 (i.e., the HOMO) and 193 (i.e., the LUMO), both of which are highlighted in bold.

| Surface Energy Sensitivity | | | | | Adsorption Energy Sensitivity | | | | |
|---|---|---|---|---|---|---|---|---|---|
| Band Index | $Ti_s(3d)$ (%) | $Ti_u(3d)$ (%) | $O(2p)$ (%) | $O(2p_z)$ (%) | Band Index | $Ti_s(3d)$ (%) | $Ti_u(3d)$ (%) | $O(2p)$ (%) | $O(2p_z)$ (%) |
| 181 | 0 | 0 | 0 | 0 | **192** | **0.1** | **0.1** | **4** | **1.2** |
| 182 | 0.2 | 0.2 | 3 | 0 | **193** | **1.3** | **1.3** | **0.8** | **0.8** |
| 183 | 0 | 0 | 27 | 0 | 194 | 0.8 | 0.8 | 0.4 | 0.4 |
| 184 | 0.3 | 0.3 | 6.2 | 6.2 | 195 | 9.7 | 9.7 | 1.8 | 0 |
| 185 | 0 | 0 | 7.6 | 2.4 | 196 | 13.4 | 13.4 | 1.6 | 0 |
| 186 | 0 | 0 | 10 | 0 | 197 | 17.1 | 17.1 | 2 | 0 |
| 187 | 0.1 | 0.1 | 15.4 | 0 | 198 | 7.7 | 7.7 | 1.2 | 0 |
| 188 | 0.2 | 0.2 | 7.2 | 0 | 199 | 1.3 | 1.3 | 0 | 0 |
| 189 | 0.7 | 0.7 | 14.4 | 10.4 | 200 | 0 | 0 | 0 | 0 |
| 190 | 0.2 | 0.2 | 47.4 | 8.8 | 201 | 0 | 0 | 0 | 0 |
| 191 | 0 | 0 | 35.8 | 2.4 | 202 | 15.2 | 15.2 | 2.4 | 0 |
| **192** | **0.1** | **0.1** | **4** | **1.2** | 203 | 0 | 0 | 0 | 0 |
| **193** | **1.3** | **1.3** | **0.8** | **0.8** | 204 | 4.7 | 4.7 | 1.8 | 1.2 |
| 194 | 0.8 | 0.8 | 0.4 | 0.4 | 205 | 0 | 0 | 0 | 0 |
| 195 | 9.7 | 9.7 | 1.8 | 0 | 206 | 0.9 | 0.9 | 1.6 | 0 |
| 196 | 13.4 | 13.4 | 1.6 | 0 | 207 | 18.7 | 18.7 | 2.8 | 0 |
| 197 | 17.1 | 17.1 | 2 | 0 | 208 | 17.7 | 17.7 | 2.8 | 0 |
| 198 | 7.7 | 7.7 | 1.2 | 0 | 209 | 0 | 0 | 0 | 0 |
| 199 | 1.3 | 1.3 | 0 | 0 | 210 | 10.5 | 10.5 | 3.6 | 2.8 |
| 200 | 0 | 0 | 0 | 0 | 211 | 0 | 0 | 0 | 0 |

**Table S19.** Atomic orbital (AO) contributions from the Ti$_s$(3$d$), Ti$_u$(3$d$), O*(2$p$) and top surface layer O(2$p$) orbitals for the fractionality-based best selected plane-wave states (i.e., "molecular orbitals", labeled by their band index) for Wannier function construction in O* decorated TiO$_2$(110) for improving adsorption energy sensitivity. All AO contributions are computed at the Γ-point using their projections on the AOs and are given as a percentage. The Ti(3$d$) and O(2$p$) columns contain the sum of contributions from all the individual $d$ and $p$ AOs respectively. Contributions from O(2$s$) AOs, Ti$_s$(3$s$/3$p$/4$s$) AOs, Ti$_u$(3$s$/3$p$/4$s$) AOs or from the AOs of any other Ti or O atoms in the surface model are not shown. Any other missing contribution (i.e., rows that sum up to less than 100%) was not suitably described by the atomic projections available in the pseudopotentials being used. The Fermi level lies between the bands indexed 194 (i.e., the HOMO) and 195 (i.e., the LUMO), both of which are highlighted in bold.

| Band Index | Ti$_s$(3$d$) (%) | Ti$_u$(3$d$) (%) | O(2$p$) (%) | O*(2$p$) (%) |
|---|---|---|---|---|
| 189 | 0.9 | 0.5 | 29.1 | 1.2 |
| 190 | 0.1 | 0.3 | 0 | 1.3 |
| 191 | 0 | 0 | 0 | 0 |
| 192 | 0 | 0 | 0 | 0 |
| 193 | 4.7 | 2.3 | 28.4 | 51.9 |
| **194** | **0** | **0** | **0.7** | **0.2** |
| **195** | **0** | **0.1** | **47.2** | **7.5** |
| 196 | 0.2 | 0.7 | 33.4 | 7.8 |
| 197 | 0 | 0 | 0 | 0 |
| 198 | 0 | 0 | 0 | 0 |
| 199 | 0 | 0 | 0 | 0 |
| 200 | 0 | 0 | 0 | 0 |
| 201 | 0 | 0 | 0 | 0 |
| 202 | 0 | 0 | 0 | 0 |
| 203 | 0 | 0 | 0 | 0 |
| 204 | 0 | 0.8 | 0 | 0 |
| 205 | 0 | 0 | 0 | 0 |
| 206 | 0.4 | 4.2 | 0 | 0.5 |
| 207 | 0.1 | 3.8 | 0.1 | 0 |
| 208 | 0 | 0 | 0 | 0 |

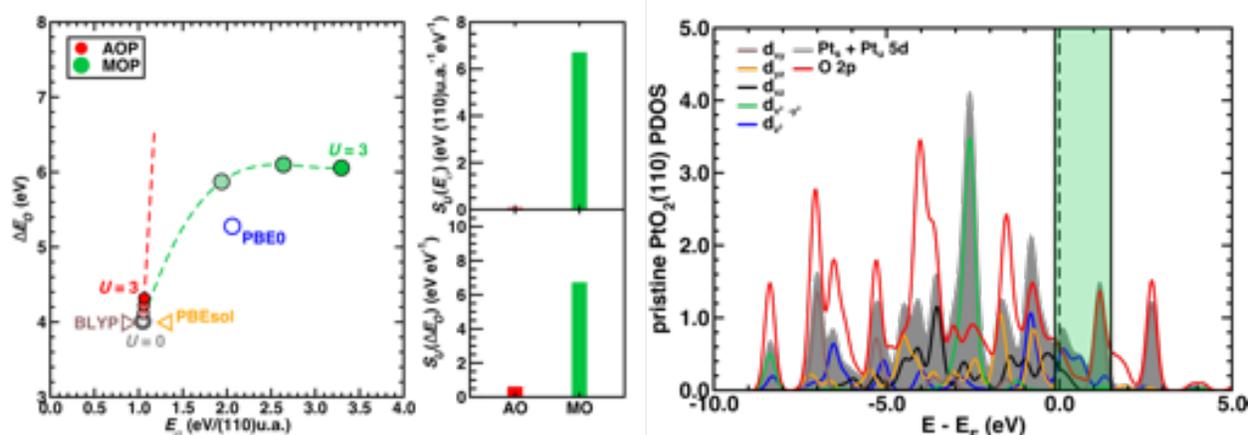

**Figure S16.** (left) $\Delta E_O$ (in eV) versus $E_\sigma$ (in eV/(110)u.a.) for $PtO_2$ (110) computed using Pt $5d$ AOPs (shaded red circles) and MOPs (shaded green circles) at integer values of $U$ from 0 to 3 eV with a dashed line indicating a fit between these points. Values from BLYP[8,9] (brown triangle), PBEsol[10] (orange triangle), and PBE0[4,6] (blue circle) were obtained from Ref. 1 using a localized basis set and shifted to align our PBE plane-wave basis set value and the PBE localized basis set value from Ref. 1. Fractionality-selected best MOPs were used for the bulk and O* decorated $PtO_2$ models and MOPs were chosen for the pristine $PtO_2$ slab to be identical in both surface formation and adsorption calculations. (middle) Linearized sensitivities of $E_\sigma$ (top, in eV/(110) unit area per eV of $U$) and of $\Delta E_O$ (bottom, in eV per eV of $U$) computed with AOPs (red bars) and MOPs (green bars). (right) Γ-point projected density of states (PDOS) of all $Pt_s(5d)$ and $Pt_u(5d)$ atomic orbitals (grey shaded region) and of the coordinating $O(2p)$ atomic orbitals in the (110) plane (red solid line) for pristine $PtO_2(110)$. The PDOS for the Pt $5d_{xy}$, $5d_{yz}$, $5d_{xz}$, $5d_{x^2-y^2}$, and $5d_{z^2}$ atomic orbitals are shown in brown, orange, black, green, and blue, respectively. The vertical black dashed line indicates the Fermi level ($E_f$). The green transparent region represents the chosen alternate states used for constructing the molecular projectors within the Wannier localization scheme for improving both surface energy and adsorption energy sensitivities by maximizing the geometric mean in $\Delta Tr[\mathbf{n}(1-\mathbf{n})]$ for both quantities.

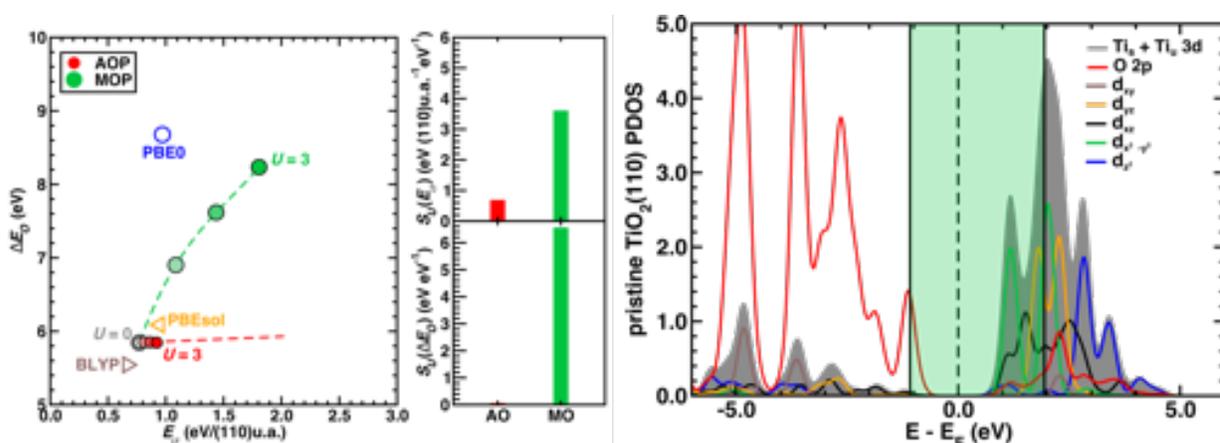

**Figure S17.** (left) $\Delta E_O$ (in eV) versus $E_\sigma$ (in eV/(110)u.a.) for TiO$_2$ (110) computed using Ti $3d$ AOPs (shaded red circles) and MOPs (shaded green circles) at integer values of $U$ from 0 to 3 eV with a dashed line indicating the trend between these points. Values from BLYP[8,9] (brown triangle), PBEsol[10] (orange triangle), and PBE0[4,6] (blue circle) were obtained from Ref. 1 using a localized basis set and shifted to align our PBE plane-wave basis set value and the PBE localized basis set value from Ref. 1. Fractionality-selected best MOPs were used for the bulk and O* decorated TiO$_2$ models and MOPs were chosen for the pristine PtO$_2$ slab to be identical in both surface formation and adsorption calculations. (middle) Linearized sensitivities of $E_\sigma$ (top, in eV/(110) unit area per eV of $U$) and of $\Delta E_O$ (bottom, in eV per eV of $U$) computed with AOPs (red bars) and MOPs (green bars). (right) Γ-point projected density of states (PDOS) of all Ti$_s$($3d$) and Ti$_u$($3d$) atomic orbitals (grey shaded region) and of the coordinating O($2p$) atomic orbitals in the (110) plane (red solid line) for pristine TiO$_2$(110). The PDOS for the Ti $3d_{xy}$, $3d_{yz}$, $3d_{xz}$, $3d_{x^2-y^2}$, and $3d_{z^2}$ atomic orbitals are indicated in brown, orange, black, green, and blue, respectively. The vertical black dashed line denotes the Fermi level (E$_f$). The green transparent region represents the chosen alternate states used for constructing the molecular projectors within the Wannier localization scheme for improving both surface energy and adsorption energy sensitivities by maximizing the geometric mean in $\Delta \text{Tr}[\mathbf{n}(1-\mathbf{n})]$ for both quantities.